\documentclass[twocolumn,secnumarabic,amssymb,amsmath,footinbib,tightenlines,nobibnotes,aps,prb]{revtex4}
\usepackage{pict2e}
\usepackage{makecell}
\usepackage{amsfonts}
\usepackage{mathtools}
\usepackage[usenames,dvipsnames]{color}
\usepackage{hyperref}
\usepackage{graphicx}
\usepackage{color}
\usepackage{subfigure}

\newcommand\nd{^{\vphantom{\dagger}}}
\newcommand\yd{^\dagger}

\newcommand{\mm}[1]{\mathcal{M}_{#1}}
\newcommand{\mme}[1]{\left[\mm{#1}\right]}
\newcommand{\qq}{\mathcal{Q}}
\newcommand{\qqe}[1]{\left[\mathcal{Q}_{#1}\right]}
\def\frac#1#2{{\textstyle{#1 \over #2}}}
\def\half{\frac{1}{2}}
\def\HH{{\hat H}}
\def\HK{{\hat K}}
\def\CG{{\cal G}}
\def\kL{k_x^{\scriptscriptstyle{\rm L}}}
\def\kU{k_x^{\scriptscriptstyle{\rm U}}}
\def\vrh{\varrho}
\def\vrht{{\tilde\varrho}}
\def\EF{E_\textsf{F}}
\def\Cocc{C_{\rm occ}}

\mathchardef\Gamma="7100
\mathchardef\Lambda="7103

\def\wwp{{\raise2.5pt\hbox{$\wp$}}}

\begin{document}

\title{Entanglement spectrum and Wannier center flow of the Hofstadter problem}
\author{Zhoushen Huang}
\author{Daniel P. Arovas}
\affiliation{Department of Physics, University of California at San Diego, La Jolla CA 92093}
\date{\today}

\begin{abstract}
We examine the quantum entanglement spectra and Wannier functions of the square lattice Hofstadter model.
Consistent with previous work on entanglement spectra of topological band structures, we find that the
entanglement levels exhibit a spectral flow similar to that of the full system's energy spectrum.  While the energy
spectra are continuous, with cylindrical boundary conditions the entanglement spectra exhibit discontinuities
associated with the passage of an energy edge state through the Fermi level.  We show how the entanglement
spectrum can be understood by examining the
band projectors of the full system and their behavior under adiabatic pumping.  In so doing we make connections
with the original TKNN work\cite{tknn82} on topological two-dimensional band structures and their Chern numbers.
Finally we consider Wannier states and their adiabatic flows, and draw connections to the entanglement properties.

\end{abstract}


\maketitle

\renewcommand{\mod}{\text{mod}}
\section{Introduction}
In the presence of a uniform magnetic field, the energy spectrum of a noninteracting two-dimensional electron gas
is arranged in discrete, equally spaced Landau levels.   The Hall conductivity of $n$ filled Landau levels is $\sigma_{xy}=ne^2/h$.
A discretized version of this model, due to Hofstadter\cite{hof76}, has electrons hopping on a two-dimensional lattice with
complex tight-binding amplitudes $t_{ij}=t\,e^{iA_{ij}}$, such that the magnetic flux through each unit cell is a rational multiple
$p/q$ of the Dirac flux quantum.  The gauge field $A_{ij}$ can be made periodic by choosing a magnetic unit cell comprising
$q$ structural unit cells of the lattice.  For a lattice with an $r$ element basis, this results in $qr$ energy subbands which in general
do not cross, a consequence of the Wigner-von Neumann theorem; the continuum limit is recovered at low energies for $q\to\infty$.
Plotting these energies as a function of $\phi\equiv 2\pi  p/q$ yields the famous `Hofstadter butterfly'.   As shown in a seminal paper
by Thouless {\it et al.\/}\cite{tknn82} (TKNN), to each band index $j$ there corresponds an integer Chern number $C_j$, which
physically represents the contribution to the Hall conductivity when band $j$ is filled.  The main differences with respect to the continuum
are (i) the tight-binding subbands are dispersive, and (ii) whereas $C_j=1$ for each Landau level in the continuum, the Chern indices of
the TKNN bands are in general nonuniform.   

The Chern number is an integer invariant which reflects aspects of the bulk band topology.  As such it is robust and invariant with respect
to parameter variations which do not collapse the band gaps.  The nontrivial bulk topology is also manifested at the edge.  Hatsugai \cite{hatsugai93}
showed that the number of edge modes interpolating between bulk bands separated by a gap is equal to the sum of the Chern indices
of all bands below that gap.  The spectral flow of the edge energy levels as a function of the momentum parallel to the edge is also reflected
in the behavior of the quantum entanglement spectrum \cite{Li-Haldane08,haldane09} of the many-body reduced density matrix obtained
by partitioning the system along a translationally-invariant boundary.  For noninteracting fermions, the spectrum of the reduced density matrix
itself corresponds to that of a noninteracting `entanglement Hamiltonian' determined by the one-body correlation matrix of the original system
\cite{cheong04a,cheong04b,peschel03,peschel04}.  However there are also exceptions to the edge-entanglement correspondence.
For example, the entanglement
spectrum has protected midgap modes for a system with inversion symmetry even if the edge modes are
gapped\cite{turner10-inversion,hughes-prodan-bernevig11-inversion}.  In certain cases, one also has to tune the boundary conditions for a
system with nontrivial topology in order for its energy edge modes to be gapless\cite{qi-wu-zhang06-tbc}, while such tuning is not required to
observe the entanglement spectral flow.  Thus in certain sense, the entanglement spectrum is a more robust test of the bulk topology.

Once one specifies the wavevector ${\vec k}_\perp$ along the translationally invariant partition boundary, the entanglement Hamiltonian
becomes effectively one-dimensional, and the localization properties of such states can be considered from the perspective of a Wannier
basis \cite{kohn59-wannier}, and several recent studies of topological insulators have invoked Wannier states
\cite{soluyanov-vanderbilt11-z2-wannier,Yu11-z2-wannier} in their analyses.  While nonvanishing Chern numbers provide an obstruction
which rules out exponentially localized Wannier states in higher dimensions \cite{brouder07-wannier-obstruction}, the one-dimensional
entanglement eigenstates at fixed ${\vec k}_\perp$ can be so localized, and for topologically nontrivial bulk band topologies, their Wannier
centers exhibit a spectral flow similar to that observed in the edge and entanglement spectra.

In this paper, we investigate the spectral flow of entanglement levels and Wannier states derived from the two-dimensional square lattice
Hofstadter model.  We identify the correspondence between these flows and the Chern numbers of the bulk bands,
and investigate the effect of adiabatic pumping on the wavefunctions of different energy bands of the system.
We consider both energy eigenstates as well as `entanglement eigenstates' of the corresponding reduced density matrix
which results from a spatial partitioning of the system into two parts.

\section{Hofstadter model and its entanglement spectrum}

\subsection{Hofstadter model}
The Hofstadter model\cite{hof76} is a discrete model of electrons in two space dimensions and in the presence
of magnetic flux.  It is defined by a lattice tight-binding Hamiltonian,
\begin{gather}
\HH = - \sum_{\langle ij\rangle} \Big[ t\nd_{ij} \, c\yd_i c\nd_j + {\rm H.c.} \Big] \ ,
\end{gather}
where $t_{ij} = |t_{ij}|\,\,e^{iA_{ij}}$ is the complex hopping amplitude between sites $i$ and $j$.
The $\textsf{U}(1)$ flux $\phi_p$ through a plaquette $p$ is the product $\prod_{\partial p} e^{iA_{ij}}$
over a counterclockwise path of links along its boundary, $\partial p$.  We shall only consider the case of uniform
amplitude hopping, {\it  i.e.\/} $|t_{ij}|=1$.

In the continuum limit, energy eigenstates of ballistic electrons collapse into macroscopically degenerate,
equally spaced Landau levels.  The degeneracy of each Landau level is $N\nd_{\rm L}=B\Omega/\phi\nd_0$, where $B$ is
the magnetic field, $\Omega$ the total area covered by the system, and $\phi\nd_0=hc/e$ is the Dirac flux quantum.
The spectral flow of entanglement eigenstates in this limit was investigated by Rodr{\'\i}guez and Sierra\cite{Rodriguez09}.
On the lattice, the Landau levels are no longer degenerate, but form magnetic subbands, each subband accommodating 
$N\nd_{\rm L}$ states.  The model may be defined on any lattice, but for definiteness we consider the square lattice.
Our principal results do not depend qualitatively on the underlying lattice structure. (See Appendix \ref{triangle} for the case of triangular lattice.)

As is well-known, while $\phi_p$ is periodic on the scale of the structural unit cell, the vector potential
$A_{ij}$ is not.  However, if the flux $\phi$ per plaquette is uniform and is $2\pi$ times a rational number $p/q$,
a gauge can be chosen where $A_{ij}$ is periodic on the scale of a `magnetic unit cell' comprising $q$
elementary structural cells.  For example, one can choose
\begin{equation}
A_{ij}=\phi\,y_i\,\delta_{x_i,x_j+1}\,\delta_{y_i,y_j}\ ,
\end{equation}
where $(x_i,y_i)$ are integer coordinates for lattice site $i$.   The magnetic unit cell is then a $1\times q$ tower of lattice cells, and
one obtains a $q\times q$ Hamiltonian matrix with nonzero matrix elements $H_{n,n}=-2\cos(k_x+n\phi)$, $H_{n,n+1}=-1$, 
$H_{N,1}=-e^{ik_y}$, and remaining elements determined by hermiticity.  This results in $q$ magnetic subbands with dispersion
$\varepsilon_a(k_x,k_y)$.  Here we are concerned with entanglement spectra, and to this end we consider a cylinder with periodic
boundary conditions in the $x$-direction and $N_y$ sites in the $y$-direction.  The Hamiltonian matrix is 
\begin{widetext}
\begin{gather}
\label{hkx}
H(k_x, N_y, z) = -\begin{pmatrix}
2\cos(k_x + \phi) & 1 & 0 &  \cdots & z^{*}\\
1 & 2\cos(k_x + 2\phi) & 1 & & 0\\
0 & 1 & \ddots &  & \vdots \\
\vdots & & & & 1\\
z & 0 &  \cdots & 1 & 2\cos(k_x + N_y\phi)
\end{pmatrix}\ .
\end{gather}
\end{widetext}
Here, $k_x$ is the Bloch phase in the $x$ direction, $N_y$ is the
number of lattice sites in the $y$ direction, and $z$ controls the
boundary condition on $y$: $z=0$ for cylindrical boundary conditions ({\it i.e.\/} periodic in the $x$-direction and open in the $y$-direction),
$z=1$ with $N_y\>{\rm mod}\>q=0$ for periodic boundary conditions; unimodular complex $z$ can be interpreted as flux threading the compactified
cylinder. Note that for $z\in\mathbb{R}$ the Hamiltonian is real and hence the eigenfunctions may be chosen to be real as well.

In what follows, we focus on the case with integer number $K_y$ of magnetic unit cells in the $y$ direction ($N_y = q K_y $),
which is more convenient for switching between open and periodic boundary.  In ref.~\onlinecite{hatsugai93},
the edge solution is derived with the requirement of commensurability, {\it i.e.\/}, $N_y = q K_y  - 1$,
in order to exploit some structure in the transfer matrix formalism.  This restriction can be lifted in the thermodynamic limit
$N_y \rightarrow \infty$, where the spectrum of edge states localized at $y=1$ is unchanged, while that of states localized at $y=N_y$ is
shifted in $k_x$.  A proof is given in Appendix \ref{incommensurate}.

\begin{figure}[!t]
\begin{center}
\includegraphics[width=8cm,height=12cm,angle=0]{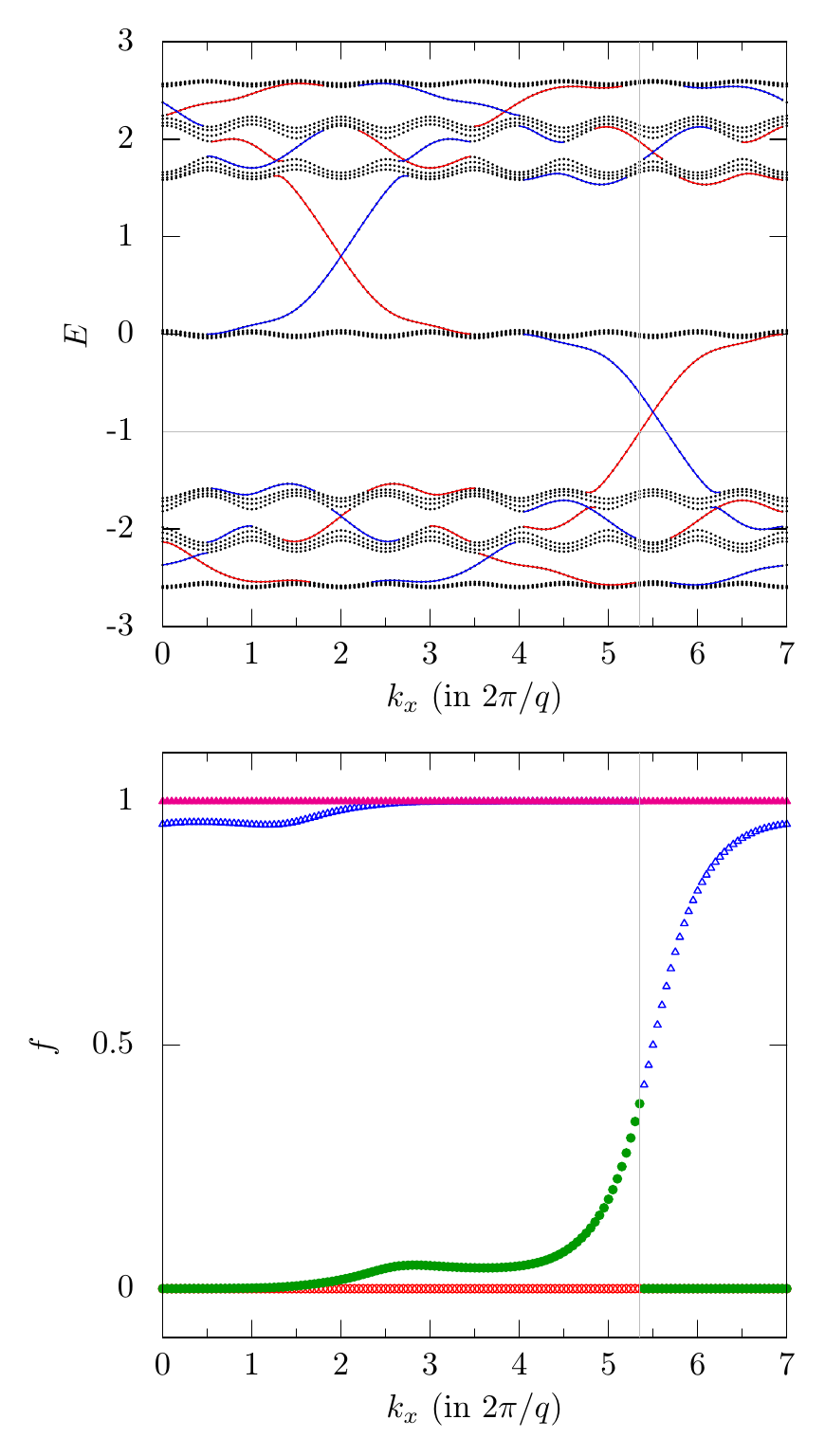}
  \caption{(Color online) Energy levels (top) and entanglement occupancies (bottom) for the square
  lattice Hofstadter model with flux $p/q=3/7$ per plaquette, on a cylinder of height $N_y=28$
  as a function of the conserved crystal momentum $k_x$.
  The Fermi level $\EF$ lies inside the third gap (grey line), and the occupied bands below $\EF$
  contribute a total Chern number $\Cocc=1$.  The energy spectrum (top) is shown for cylindrical boundary
  conditions, with the black dots indicating bulk levels, the red lines indicating edge levels localized along
  the lower edge ($y=1$), and the blue lines indicating edge levels localized along the upper edge ($y=N_y=28$).
  The vertical gray line marks the value of $k_x$ where the lower edge mode crosses the Fermi level.
  The entanglement occupancies $f_a$ (bottom) are computed for the lower half of the cylinder ($1\le y \le 14$),
  color and symbol-coded according to $a$.  Although the overall flow appears continuous, there is a discontinuity
  in the occupancies $f_a(k_x)$ where the lower edge mode crosses $\EF$, resulting in a sudden color change in the plot. }
\label{p3q7ny28} 
\end{center}
\end{figure}

\begin{figure}[!t]
\begin{center}
\includegraphics[width=8cm,height=12cm,angle=0]{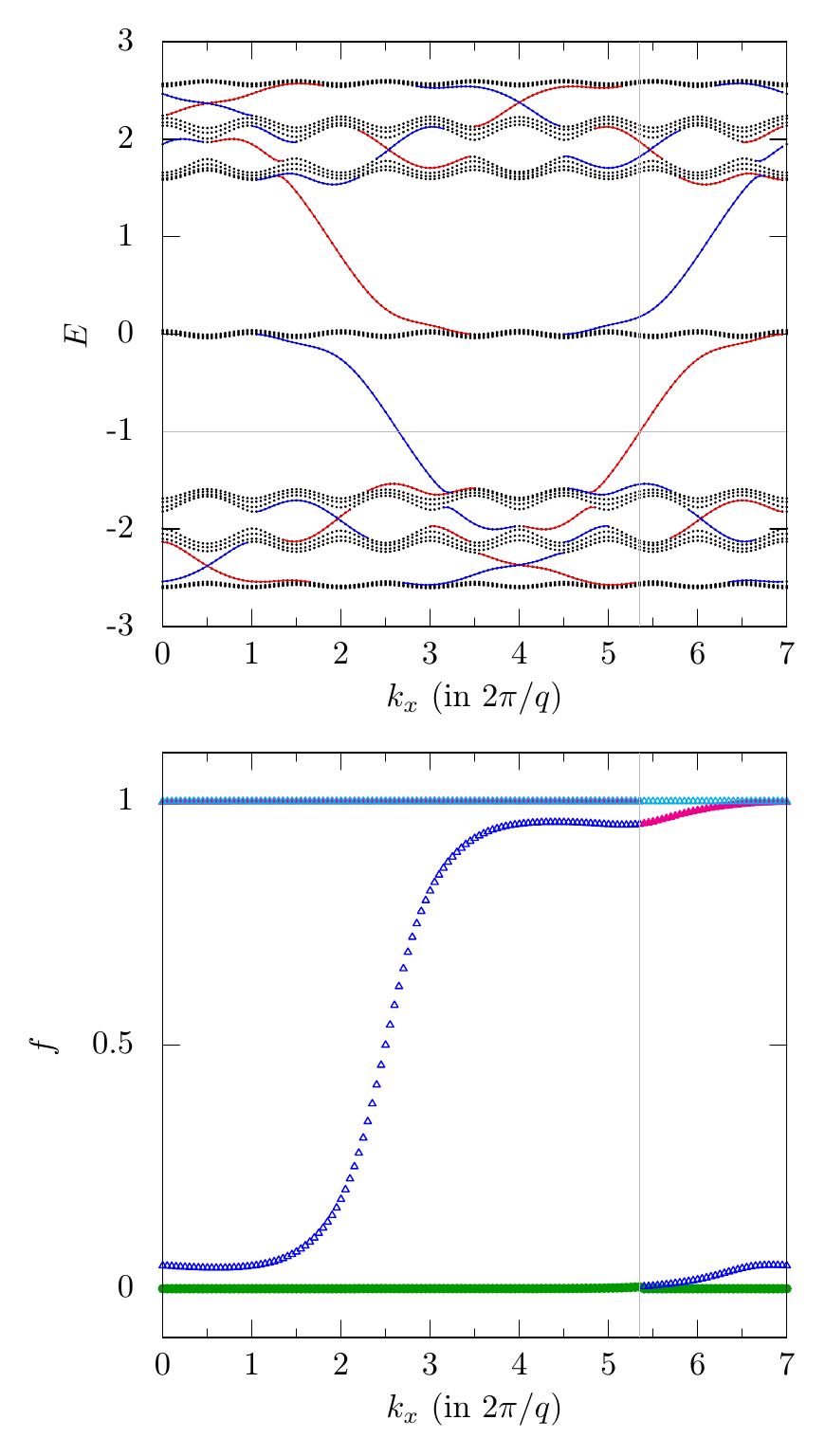}
  \caption{(Color online) Same as in Fig. \ref{p3q7ny28} but with $N_y=29$ and $M=15$.}
\label{p3q7ny29} 
\end{center}
\end{figure}

\subsection{Entanglement spectrum}
Imagine partitioning the sites of our lattice into two groups, A and B.  If a many-body wavefunction
$|\Psi\rangle$ can be written as a direct product $|\Psi^{\rm A}\rangle\otimes|\Psi^{\rm B}\rangle$, the wavefunction
is said to be unentangled with respect to this partition.  More generally, let $\varrho=|\Psi\rangle\langle\Psi|$ be the projector
onto $|\Psi\rangle$.  Tracing out over the B sites yields the reduced density matrix ${\tilde\varrho}=\textsf{Tr}\nd_{\rm B}\,\varrho$,
whose eigenvalues constitute the entanglement spectrum\cite{Li-Haldane08}.  The von Neumann entropy, $S=-\textsf{Tr}\,{\tilde\varrho}
\ln{\tilde\varrho}$, provides a measure of the degree of entanglement.  

If the boundary between A and B is irregular, then translational invariance is completely broken, but if the boundary is such that the 
A region remains periodic in one direction (or more than one, in the case of systems in more than two space dimensions),
then the eigenstates of ${\tilde\varrho}$ can be classified by a corresponding crystal momentum, and one can
investigate the spectral flow of the entanglement levels \cite{haldane09}.

\subsubsection{The correlation matrix method}
A general recipe for computing the reduced density matrix for noninteracting Fermi systems has been derived by
Cheong and Henley\cite{cheong04a}.  Let $I$ and $J$ denote sites in the full system, whose Hamiltonian is
$\HH=H\nd_{IJ}\,c\yd_I c\nd_J$.  The statistics of this Hamiltonian are then completely determined by the
one-body correlation matrix ${\cal G}\nd_{IJ}=\langle c\yd_I c\nd_J \rangle = \textsf{Tr}\, (\vrh\,c\yd_I c\nd_J)$,
where $\vrh$ is the density matrix.  Now consider a bipartition of the full system into two subsystems $A$ and $B$,
and let $i$ and $j$ denote sites within $A$.  Then $G\nd_{ij}=\langle c\yd_i c\nd_j\rangle = (R\CG R^\textsf{T})_{ij}$
where $R\nd_{iI}=\delta\nd_{iI}$ is an oblong matrix of dimensions $N_A\times(N_A+N_B)$ with $1$'s along the
diagonal; $N_{A(B)}$ is the size of the $A(B)$ subspace.  {\it I.e.\/} $R$ spatially projects onto $A$.
Thus $G$ is a submatrix of $\CG$, and a key fact, due to Peschel\cite{peschel04}, is that we may write
$G\nd_{ij}=\textsf{Tr}\, (\vrht\,c\yd_i c\nd_j)$, where 
$\vrht=\exp(-\HK)/Z$ is the reduced density matrix (RDM) and $\HK\equiv\Gamma\nd_{ij}\,c\yd_i\,c\nd_j$
is the dimensionless `entanglement Hamiltonian' (both restricted to $A$).  One then finds 
$G=\big\{\!\exp(\Gamma^\textsf{T})+1\big\}^{-1}$, and the normalization $Z$ follows from $\textsf{Tr}\,(\vrht)=1$.
The $N_A$ eigenvalues $\{\gamma_a\}$ of $\Gamma^\textsf{T}=\ln\big(G^{-1}-1\big)$ are the entanglement
`quasienergies', and the eigenvalues of $G$ are then Fermi functions of the quasienergies\cite{cheong04b}, {\it viz.\/}
\begin{gather}
\label{fermi}
f_a = {1\over\exp(\gamma_a) + 1} \ .
\end{gather}
For our system, the translation invariance along $x$ means $k_x$ is a good quantum number, and for each $k_x$, the
system can be regarded as one-dimensional.  Thus, $H_{IJ}(k_x)$ and $\CG_{IJ}(k_x)$ are of dimension $N_y$, and $I$ and $J$
label rings and not single sites.  In our study, the Fermi energy $\EF$  is always placed within some bulk gap.  For periodic boundary
conditions, $\CG$ is then a sum of projectors onto the occupied bands.  With open boundaries, there will be edge modes which
cross the Fermi level.  In either case, we take the $A$ subsystem to be the bottom part of the cylinder, with $y\in[1,M]$.
Thus $G$ is the upper left $M\times M$ block of $\CG$.

\subsubsection{Rank of $G$ and $\mathbf{1} - G$}\label{ranksec}
The eigenvalues of $G$ may contain \emph{exact} zeros or ones. The
number of zeros and ones, denoted as $D_0$ and $D_1$ respectively, are
by definition the dimensions of the kernels of $G$ and $\mathbf{1} - G$. 
If $\nu$ bulk bands are occupied with periodic boundary conditions, then the total rank of $\CG(k_x)$ is $\nu N_y/q$, since each
of the $q$ bands contains an equal number of states.  Thus if $M\ge \nu N_y/q$, the rank of $G$ will also be $\nu N_y/q$.
For $M\le \nu N_y/q$, the rank of $G$ is $M$.  Thus,
\begin{gather}
  {\rm rank}(G)={\rm min}(M,\nu N_y/q) \ ,\\
  D_0 = M - {\rm rank}(G)\ ,
\label{rankG}
\end{gather}
and similarly
\begin{gather}
  {\rm rank}(\mathbf{1} - G) = {\rm min}\Bigl\{M, N_y(1 - \nu/q)\Bigr\}\ ,\\
  D_1 = M - {\rm rank}(\mathbf{1} - G)\ ,
\end{gather}
where $N_y(1-\nu/q)$ is the rank of $\mathbf{1} - \CG$, the projector
onto unoccupied bands.

As we shall see, with cylindrical boundary conditions, the rank of $\CG$
changes with $k_x$ whenever an edge state crosses the Fermi level. As
a result, $D_0$ and/or $D_1$ are discontinuous at such $k_x$ values
for certain range of $M$. It is easy to verify that the condition for
$D_0 = D_1 = 0$ is $M \le {\rm min}(\nu/q, 1 - \nu/q) \times N_y$.

\begin{figure}[!t]
\begin{center}
  \subfigure{\label{c2-erg}\includegraphics[width=0.45\textwidth]{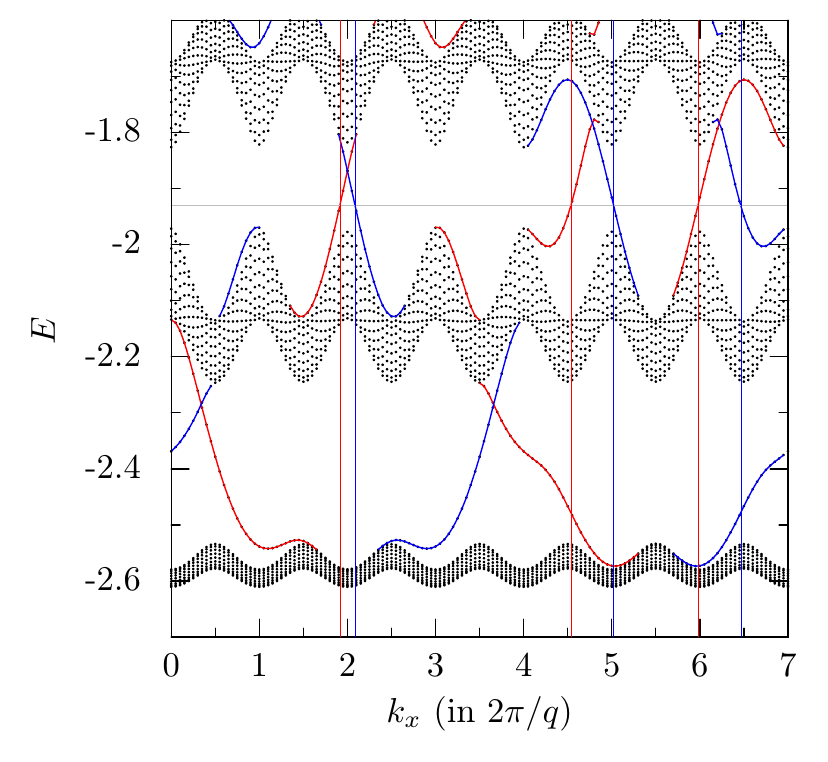}}
    \subfigure{\label{c2-f}\includegraphics[width=0.45\textwidth]{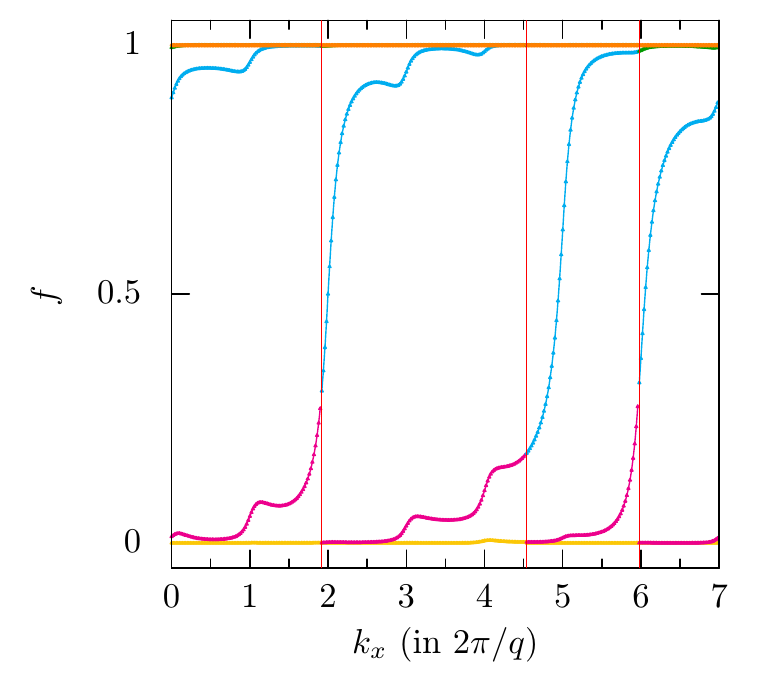}}
  \caption{(Color online) Hofstadter model with flux $p/q=3/7$ per plaquette on a cylinder of height $N_y=70$.
  Top: detail of energy spectrum showing
  lowest three energy bands.  The Fermi level lies at $\EF = -1.9$ (grey horizontal line).
  The total Chern number of the occupied bands is $\Cocc=3$.
  Red (blue) vertical lines indicate $k_x$ at which the lower (upper) edge modes cross the Fermi energy ($\kL$ and $\kU$ in text).
  Bottom: entanglement occupancies $f_a$ after tracing out the upper half of the cylinder.}
\label{c2a} 
\end{center}
\end{figure}

\begin{figure}[!t]
\begin{center}
  \subfigure{\label{c2-gamma}\includegraphics[width=0.45\textwidth]{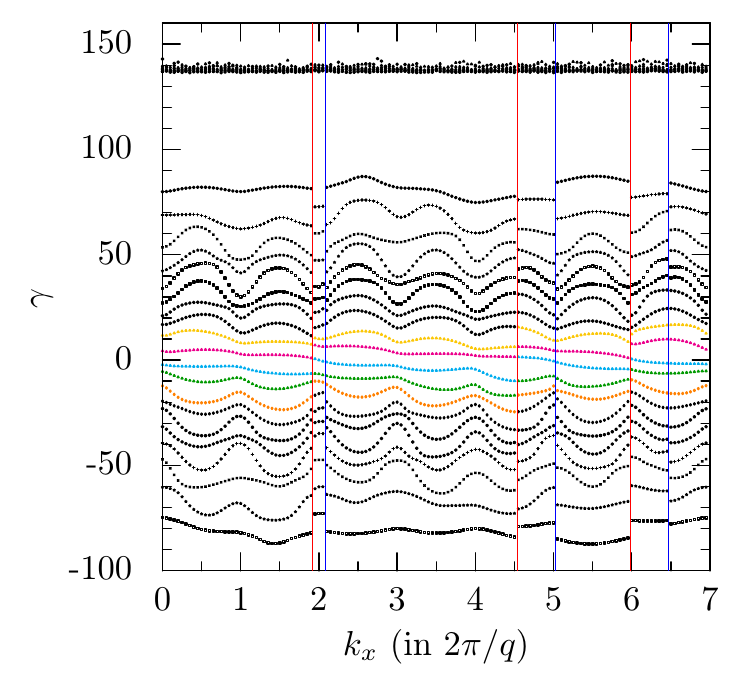}}
  \subfigure{\label{p3-q7-c2-gamma-polar-dense}\includegraphics[width=0.43\textwidth]{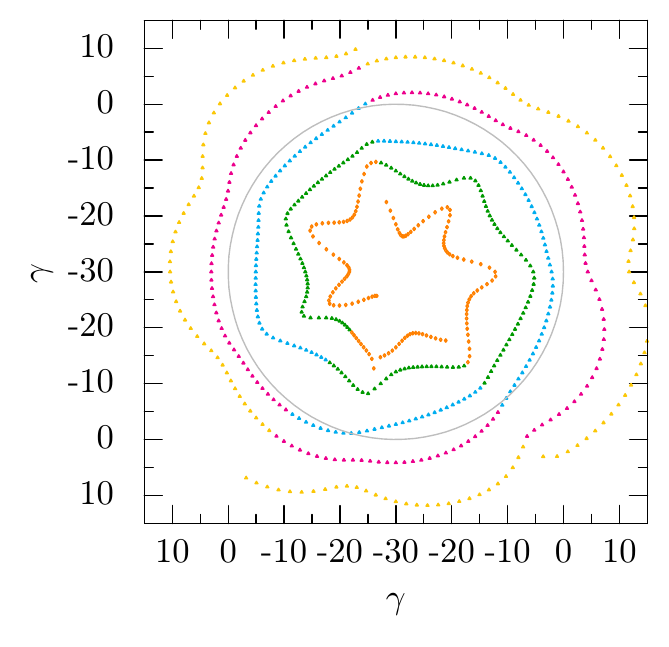}}
    \caption{(Color online) Same system as in Fig. \ref{c2a}.  Top:  entanglement quasi-energies $\gamma_a$.
   $20$ levels are below the large `gap' -- the same as the number of occupied levels in the full system.
   Colored levels are also plotted in Fig. \ref{c2a} with the same color scheme. Red and blue vertical lines mark the $k_x$ values at which lower and upper edge modes of the cylindrical Hamiltonian are crossed by the Fermi level ($\kL$ and $\kU$ in text).
    Bottom: quasienergy in 2D polar coordinates.  The radius is the quasienergy
    and the polar angle is $k_x$.  The black circle corresponds to $\gamma = 0$.  There are
    three curves spiraling outward in the clockwise direction, corresponding to total Chern number $C=3$.}
  \label{c2b}
  \end{center}
\end{figure}

\subsubsection{Entanglement occupancy, quasienergy, and spectral flow}

Fig.~\ref{p3q7ny28} shows the energy spectrum and entanglement
occupancy for $p/q = 3/7$, $N_y = 28$, and $M = 14$, using cylindrical boundary
conditions.  The Fermi energy $\EF$ lies within the third gap.
The total Chern number of the three occupied bands is $\Cocc=\sum_{j=1}^3 C_j=1$, which
is also the number of times the lower edge mode, localized at
$y=1$ (the red line in the plot), flows across the gap, {\it i.e.\/}, the
winding number\cite{hatsugai93}. The sign of $\Cocc$ is reflected in
the direction of the edge flow, {\it e.g.\/} $\Cocc = -1$ for $4$ filled
bands, and the lower edge flows downward.  Several features are noteworthy:
\smallskip

\noindent (a) Most levels are clustered near $f=0$ and $f=1$. This reflects the fact
that $\EF$ lies inside a bulk gap.  Taking linear combinations of the occupied states
in the full system, one can create wavefunctions which are mostly confined to either
(A or B) subsystem.  The same consideration applies to unoccupied states.
In the thermodynamic limit, the fraction of occupied states in A and in B
should be the same as that for the full system.  This is confirmed by our numerical results.
The entanglement eigenstates with $f\sim\frac{1}{2}$ are localized along $y\approx M$.
As is the case for the edge modes of the full Hamiltonian, the number of entanglement levels flowing
between $f\simeq 0$ and $f\simeq1$ is the same as the total Chern number number $\Cocc$ of the filled bands. 
This is depicted in Fig. \ref{p3q7ny28} for $\Cocc=1$ and in Fig. \ref{c2-f} for $\Cocc=3$.

\smallskip

\noindent (b) The occupancy $f_a(k_x)$ is discontinuous at $k\nd_x=\kL$, where the lower cylindrical boundary edge modes
of $\HH$ cross the Fermi level; these are the red curves in Fig. \ref{c2-erg}.  For example, in Figs.~\ref{p3q7ny28} and \ref{p3q7ny29},
where entanglement levels with different indices $a$ are plotted in different colors, the tenth (light blue) and eleventh
(magenta) level occupancies are each discontinuous at $\kL$, but satisfy $f_{a=10}(\kL+0^+)=f_{a=11}(\kL-0^+)$.
The number of distinct such $\kL$ values is the number of lower edge mode Fermi level crossings, which is $|\Cocc|$.
Thus, as one increases $k_x$ through each such crossing, the number of levels with $f\simeq 1$, and hence the total
occupancy $\sum_a f_a$ of the $A$ subsystem, drops discontinuously by
unity due to the exclusion of the edge mode.
Eventually the $f\simeq 1$ levels are repopulated due to the aforementioned spectral flow.
In order to conserve the rank of $G(k_x)$ (in the case of $M >
{\rm rank}(\CG)$, \emph{cf.}~eqn.~\ref{rankG}) upon increasing $k_x$ by $2\pi$, then, there must be a discontinuous repopulation of
the $f\simeq 0$ levels.  This occurs when the upper cylindrical boundary edge modes cross $\EF$ at $k_x = \kU$; these are the blue curves in Fig. \ref{c2-erg}.
As these modes have a vanishingly small projection onto the $A$ subsystem in the thermodynamic limit, they lead to no discontinuity 
in the total occupancy of $A$.  A similar analysis of the occupancy discontinuity has recently been given by Alexandradinata {\it et al.\/} 
\cite{AHB11}

\smallskip

\noindent (c) From the occupancy spectrum, one can invert the Fermi distribution
(eqn.~\ref{fermi}) to get the quasienergy spectrum $\{\gamma_a\}$.  A quasienergy plot more clearly
reveals entanglement spectrum near $f=0$ and $f=1$, where many levels are clustered.
In Fig. \ref{c2-gamma} (and equivalently the polar plot Fig. \ref{p3-q7-c2-gamma-polar-dense}), some
key features are apparent.   First, a substantial number of levels are clustered at $\gamma\approx 140$ ($f\sim 10^{-60}$)
and are separated from the remaining levels by a pronounced gap.   Actually this is a numerical artifact and these levels
all lie at $\gamma=\infty$.  Recall the earlier result ${\rm rank}(G)={\rm min}(M,\nu N_y/q)$ in eqn. \ref{rankG} for a system with
periodic boundary conditions.  Here we have $p/q=3/7$, $N_y=70$, $M=35$, and $\nu=2$ since $\EF$ is placed in the gap
between the second and third bulk bands.  Thus we would expect ${\rm rank}(G)=20$, and since the row dimension of $G$ is
$M=35$, there should be $15$ levels with $f=0$, corresponding to $\gamma=+\infty$. Had one looked at a system with more than half filling, one would find entanglement quasienergies clustering at $\gamma = -\infty$ instead, where the entanglement occupancy is exactly $1$. These would correspond to the kernel of $1 - G$.

\smallskip

\noindent (d) Since our system has cylindrical boundary conditions, there are edge states, and there is a discontinuity in the quasienergy
spectrum at each $\kL$ and $\kU$ value where lower and upper edge states are crossed by $\EF$.  When both edge states lie below $\EF$, one counts $20$ finite quasienergy levels.
When $k_x$ lies between consecutive $\kU$ and $\kL$ values, one of the lower boundary edge states has crossed the Fermi level,
and the rank of $G$ decreases to $19$.  The spectral flow in the vicinity of $\gamma\approx 0$ is continuous, however.   Discontinuities
in the entanglement energies occur for large values of $|\gamma|$, where the occupancy is close to $0$ of $1$.   When an edge state
passes from below $\EF$ to above $\EF$, the rank of $\CG$ changes discontinuously by $-1$.  For increasing $k_x$, this occurs at any 
of the three $\kL$ points in Fig. \ref{c2-erg}.   Such an edge state has almost perfect projection onto the $A$ subsystem, hence its
depopulation leads to a sudden rearrangement of entanglement levels with large negative quasienergies $\gamma_a$ ($f_a\approx 1$)
and a loss of one such level.  For $k_x=\kU$, where the change $\Delta\,{\rm rank}(\CG)=+1$, the `extra' level enters via a discontinuous
rearrangement of the levels with large positive quasienergies ($f_a\approx 0$).  (The situation is reversed if $\EF$ lies within the first gap,
in which case $\Delta\,{\rm rank}(\CG)=+1$ at each $\kL$ and $\Delta\,{\rm rank}(\CG)=-1$ at each $\kU$.)  We see this clearly in
Fig. \ref{c2-gamma}, where the number of finite $\gamma$ levels changes from $20$ to $19$ when $k_x$ lies between consecutive
$\kL$ and $\kU$ values.

\begin{figure}[!t]
\begin{center}
\includegraphics[width=8.7cm,angle=0]{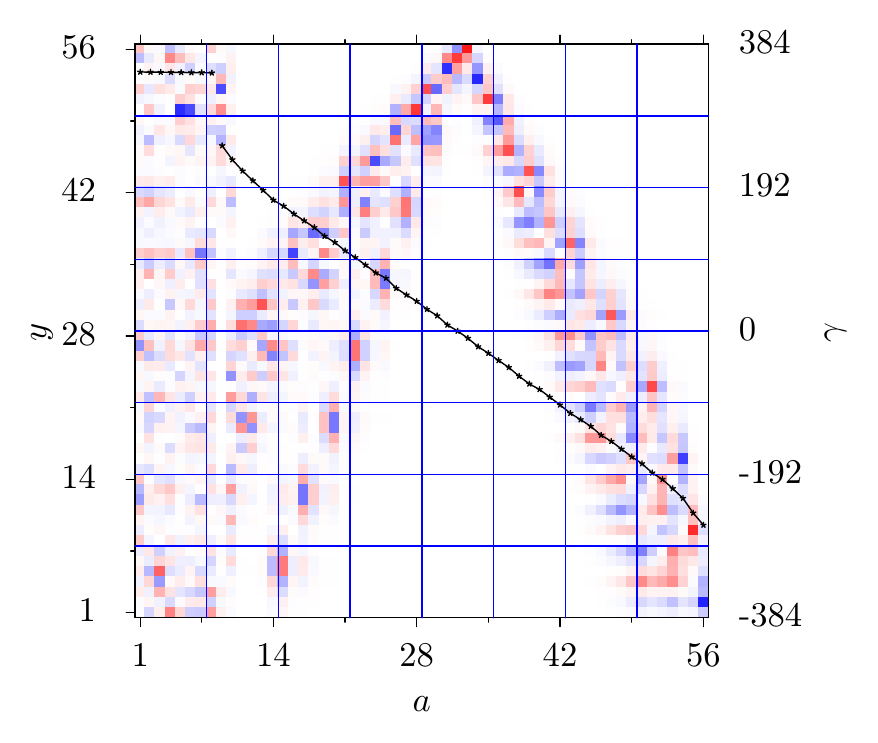}
  \caption{(Color online) Normalized eigenfunctions ${\tilde\psi}_a(y)$ of $G$ for $p/q=3/7$, $N_y=112$,
  $M=56$, and $\nu=3$ at $k_x=2\pi/7$ (cylindrical boundary conditions).  The Chern number of the filled bands is $\Cocc=1$.
  Color corresponds to the sign of the wavefunction (red for positive, blue for negative), and intensity
  to amplitude (white for zero intensity).  The black points are the entanglement energies $\gamma_a$.}
\label{Gwfs} 
\end{center}
\end{figure}

\subsubsection{Entanglement eigenfunctions}\label{entwfs}
In Fig. \ref{Gwfs} we plot the eigenfunctions ${\tilde\psi}_a$ of $G(k_x=2\pi/7)$ for a larger $p/q=3/7$ system, with $N_y=112$,
$M=56$, and $\nu=3$ at $k_x=2\pi/7$.  The Chern number of the filled bands is $\Cocc=1$.  States with finite $\gamma_a$ are
spatially resolved.  The rank of $G$ is $\nu N_y/q=48$, corresponding to states \#9 through \#56 in the plot.  The dimension of
the kernel of $G$ is then $\textsf{dim}(G)-\textsf{rank}(G)=8$.  These states all have $f_a=0$, {\it i.e.\/} $\gamma_a=+\infty$, which is rendered
as the flat ceiling of the black curve in the figure.  They form the speckled region in the left of the figure.  Note that states with
$\gamma_a\approx 0$ ($f\approx{1\over 2}$) are localized near the cut $y=M$, and that those with large $\gamma_a$ are localized
away from the cut.  We shall return to this point later, toward the end of the paper, after we discuss Wannier center flows.

\subsubsection{Effect of changing $M$}

Since the magnetic unit cell is set along the $y$-direction, $N_y$ and $M$ must be chosen as integer multiples of $q$ if there
are to be an integer number of unit cells in the full system and/or the lower ($A$) subsystem, respectively.
As shown in Appendix \ref{incommensurate}, changing $N_y$ to $N_y + m$ ($m\in\mathbb{Z}$) keeps the lower edge
modes intact, but shifts the $k_x$ values for the upper edge modes by $-2\pi m p/q$.  For example, the lower edge modes 
(red lines) are the same in Figs.~\ref{p3q7ny28} and \ref{p3q7ny29}, but the upper edge modes (blue lines) in Fig.~\ref{p3q7ny29}
are shifted in $k_x$ by $-6\pi/7$ relative to those in Fig.~\ref{p3q7ny28}.

It turns out that changing $M$ affects the entanglement occupancy in the same way as changing $N_y$ would affect the edge modes.
This is shown in the bottom rows of Figs.~\ref{p3q7ny28} and \ref{p3q7ny29}.   We should mention that keeping $M$ fixed while
changing $N_y$ will not change the occupancy spectrum in any appreciable way because that only shifts the upper edge modes.
Changing $N_y$ will thus change the $\kU$ values, and consequently where the rearrangements of the $f\approx 0$ parts of the
entanglement spectrum occur, but will not affect the spectral flow for $\gamma\approx 0$.
The reason will become more clear in the next section.

\begin{figure*}[htb]
  \subfigure[$\quad B_1$]{\includegraphics[width=0.185\textwidth]{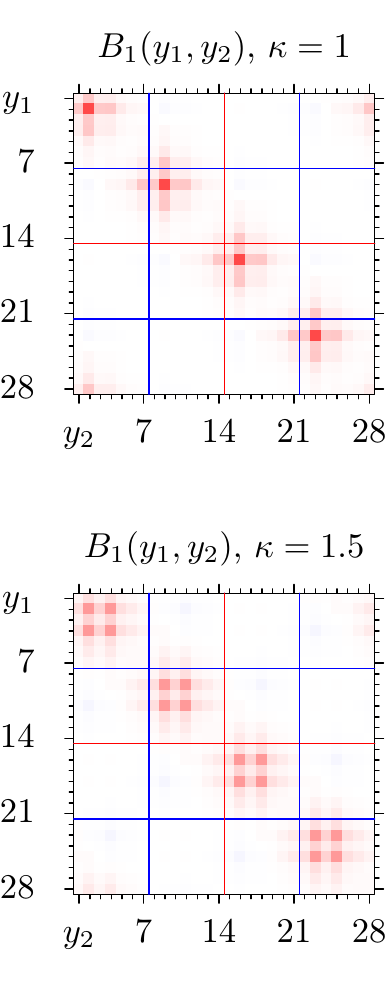}}
  \subfigure[$\quad B_2$]{\includegraphics[width=0.185\textwidth]{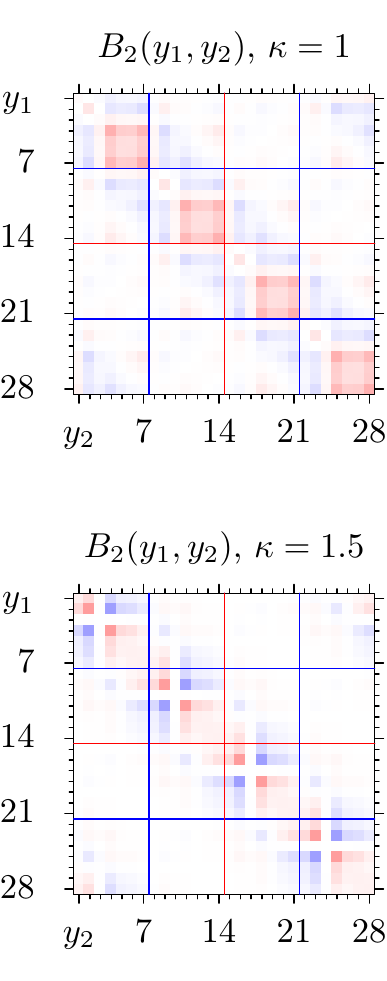}}
  \subfigure[$\quad B_3$]{\includegraphics[width=0.185\textwidth]{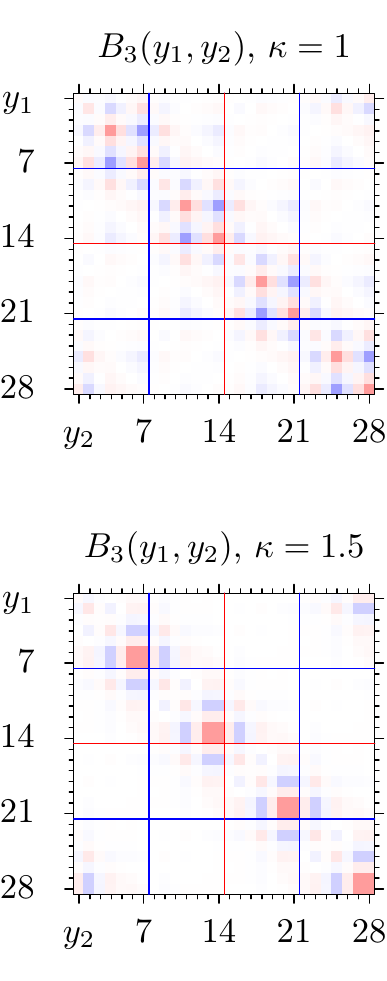}}
  \subfigure[$\quad \CG_2$]{\includegraphics[width=0.185\textwidth]{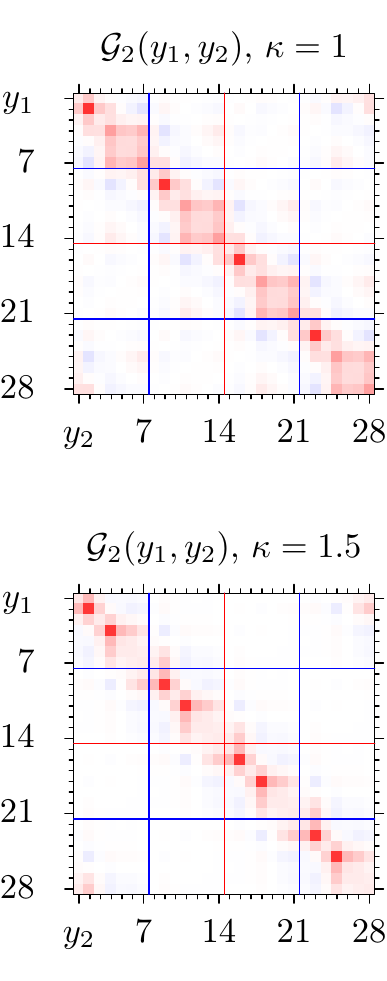}}
 \subfigure[$\quad \CG_3$]{\label{g3}\includegraphics[width=0.185\textwidth]{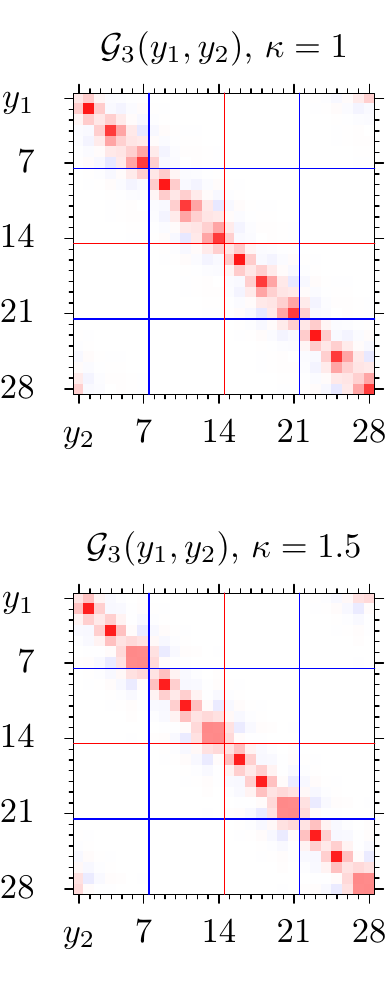}}
 \caption{(Color online) Full system band projectors for $p/q=3/7$,
   $N_y = 28$, with periodic boundary conditions in $y$ and $k_x=
   2\pi\kappa/q$. $B_j$ is the projector onto the $j^{\rm th}$ band,
   and $\CG_\nu=B_1+\ldots+B_\nu$ the projector onto the lowest $\nu$
   bands. The magnitude of the matrix elements are represented by
   intensity and their sign by color (red positive, blue negative,
   white zero). $y_1$ and $y_2$ are the row and column indices of the
   projectors. Blue rules mark boundary of the magnetic unit cells.
   Red rules mark the bipartite cut, so the top-left quadrant of $\CG$
   corresponds to the restricted correlation matrix $G$. Only $\kappa
   = 1$ and $3/2$ are shown here due to space restriction. Corresponding plots
   for other $\kappa$ values can be inferred from those shown here
   after shifting all matrix elements along the diagonal by $t$
   as the solution of eqn.~\ref{diophantine-r1}. Similarly, projectors
   of all half odd-integer $\kappa$ are obtained by shifting those of
   $\kappa = 3/2$. As $k_x$ is increased from $0$ to $2\pi$ ($\kappa$
   from $0$ to $q$), the adiabatic pumping is evident in the diagonal
   motion of all matrix elements of the projectors. Note in particular
   that at integer $\kappa$, each diagonal block of $\CG_3$ consists
   of three sharply localized packets: the top-left one is contributed by
   $B_1$, whereas the rest two result from the constructive addition
   of the diagonal blocks of $B_2$ and $B_3$. Similarly, at half
   odd-integer $\kappa$, $B_1$ and $B_2$ add constructively, yielding
   the two sharp packets in each diagonal block of $\CG_2$. The Chern
   number corresponds to how many magnetic unit cell boundaries (blue
   and red rules) any diagonal matrix element has passed by in one
   pumping period. Equivalently, it is the sum of diagonal matrix
   elements which are transferred across any magnetic unit cell
   boundary in one pumping period. For the $\CG_{\nu}$ type, it is
   intuitively how many packets are transferred. }
      \label{bigfig}
\end{figure*}

\section{Adiabatic pumping of band projectors}
\label{sec-proj}
The entanglement level occupancies $f_a$ are eigenvalues of the restricted
correlation matrix $G=R\CG R^\textsf{T}$.  In searching for an intuitive picture of
the various features of the entanglement spectrum, it is then natural to examine the
unrestricted projector $\CG$.  We found that much information can be extracted
from $\CG$ itself.

In this section, we will use periodic boundary condition in the $y$-direction, {\it i.e.\/} $z=1$ and
$N_y\,{\rm mod}\,q=0$ in eqn.~\ref{hkx}.  The Hamiltonian of eqn.~\ref{hkx} then satisfies
\begin{gather}
\label{hkx-ty}
H(k_x + \phi) = T\yd_y\,H(k_x)\,T\nd_y\ ,
\end{gather}
where $\phi=2\pi p/q$ as before and where $T_y$ is the translation operator by one lattice spacing
in the $y$-direction:
\begin{gather}
\label{ty}
T_y = \begin{pmatrix*}[l] \mathbf{0} & 1\\ \mathbf{1}\nd_{N_y - 1} & \mathbf{0} \end{pmatrix*}  \ .
\end{gather}
The unitarity of $T_y$ guarantees that the bulk bands repeat themselves
for $q$ times over the interval $k_x\in [0,2\pi]$.   With each successive increase of $k_x$ by $2\pi p/q$, the spectrum
repeats and the corresponding energy eigenstates are shifted by $\Delta y=1$.  Denoting $B_j(k_x)$ as the projector
onto the $j^{\rm th}$ band, we have that $\CG(k_x)\equiv \CG_\nu(k_x)$ is the projector onto the lowest $\nu$ filled bands,
\begin{gather}
\CG_\nu(k_x) = \sum_{j=1}^\nu B_j(k_x) \ .
\end{gather}
The covariance in $k_x$ and $y$ is reflected as
\begin{gather}
  \label{covariance}
  B_j(k_x + \phi \, ; \, y_1 \, , \, y_2) =   B_j(k_x  \, ; \, y_1+1 \, , \, y_2+1)\ ,
\end{gather}
where $y_1$ and $y_2$ are row and column indices for $B_j(k_x)$.
Translational invariance on a scale of the magnetic unit cell corresponds to
\begin{gather}
  B_j(k_x  \, ; \, y_1 \, , \, y_2) =   B_j(k_x  \, ; \, y_1+q \, , \, y_2+q)
\end{gather}
The same relations hold for $\CG_\nu$.  

While these projectors are explicitly constructed using the Bloch
states, which are spatially extended, the fact that their eigenvalues
are degenerate (either $0$ or $1$) means one may construct localized
eigenstates around the cylinder, for each $k_x$, by recombining Bloch
states of different $k_y$ with the same eigenvalue. In the continuum
limit, where $q\to\infty$ with $p$ finite, these correspond to the
familiar Landau strip basis. In fact, the projectors themselves are
localized: An illustration is provided in Fig.~\ref{bigfig}, which
shows several $B_j$ and $\CG_\nu$ for $p/q=3/7$ at $k_x=2\pi\kappa/q$
for $\kappa=1$ and $\kappa=3/2$, both of which are local extrema of
the energy bands. The size of the magnetic unit cell naturally divides
the projectors into blocks of size $q\times q$. It is not surprising
that the off-diagonal blocks drop exponentially, a consequence of the
analyticity of $B_j$ in complex $k_y$\cite{cloizeaux64.paperII}. What
is perhaps unexpected is that at band troughs (integer $\kappa$ for
odd $j$ and half-odd-integer $\kappa$ for even $j$ in
Fig.~\ref{bigfig}), the projectors $B_j$, and especially $\CG_\nu$,
are quite well-localized even within the diagonal blocks. The diagonal
matrix elements of the projectors correspond to electron density at
the corresponding $y$ coordinate. For single bands ($B_j$), any $q$
consecutive diagonal elements sum to $1$, thus one may think of them
as constituting a wavepacket, and the projector $B_j$ as consisting of
$N_y/q$ such wavepackets (one per magnetic unit cell). For each
wavepacket, the weight is dominated by one or two elements, as one can
see in Fig.~\ref{bigfig}. The localization of the projector sums
$\CG_{\nu}$ is even more prominent: when the gap between two
neighboring bands is at a minimum, their projectors add
constructively, resulting in two sharply localized dots on the
diagonal line of $\CG_\nu$. For example, in Fig.~\ref{bigfig}, $B_2$
and $B_3$ add constructively at $\kappa = 1$, yielding the lower two
dots in each diagonal block of $\CG_3$, and similarly, $B_1$ and $B_2$
add up to $\CG_2$ for $\kappa = 3/2$. Now, $q$ consecutive diagonal
elements in $\CG_{\nu}$ must sum to $\nu$, thus each of the $\nu$ dots
can be intuitively understood as one localized wavepacket. The
constructive superposition of neighboring bands then indicates the
corresponding single-band wavepackets have opposite parity so that the
off-diagonal elements cancel each other.  To relate to the
aforementioned ``strip'' states, we note that any column of a
projector is an eigenstate of the same projector, with eigenvalue
$1$\footnote{Since any projector squares to itself, each column is an
  eigenstate with eigenvalue $1$. However, they are not normalized,
  nor are they orthogonal to each other (because no eigenstate with
  eigenvalue $0$ is present). In fact the construction of Wannier
  states is one way to orthonormalize these eigenstates.}. The
diagonal nature of these band projectors thus ensures the existence of
such strip states.

\begin{figure}[!t]
\includegraphics[width=0.4\textwidth]{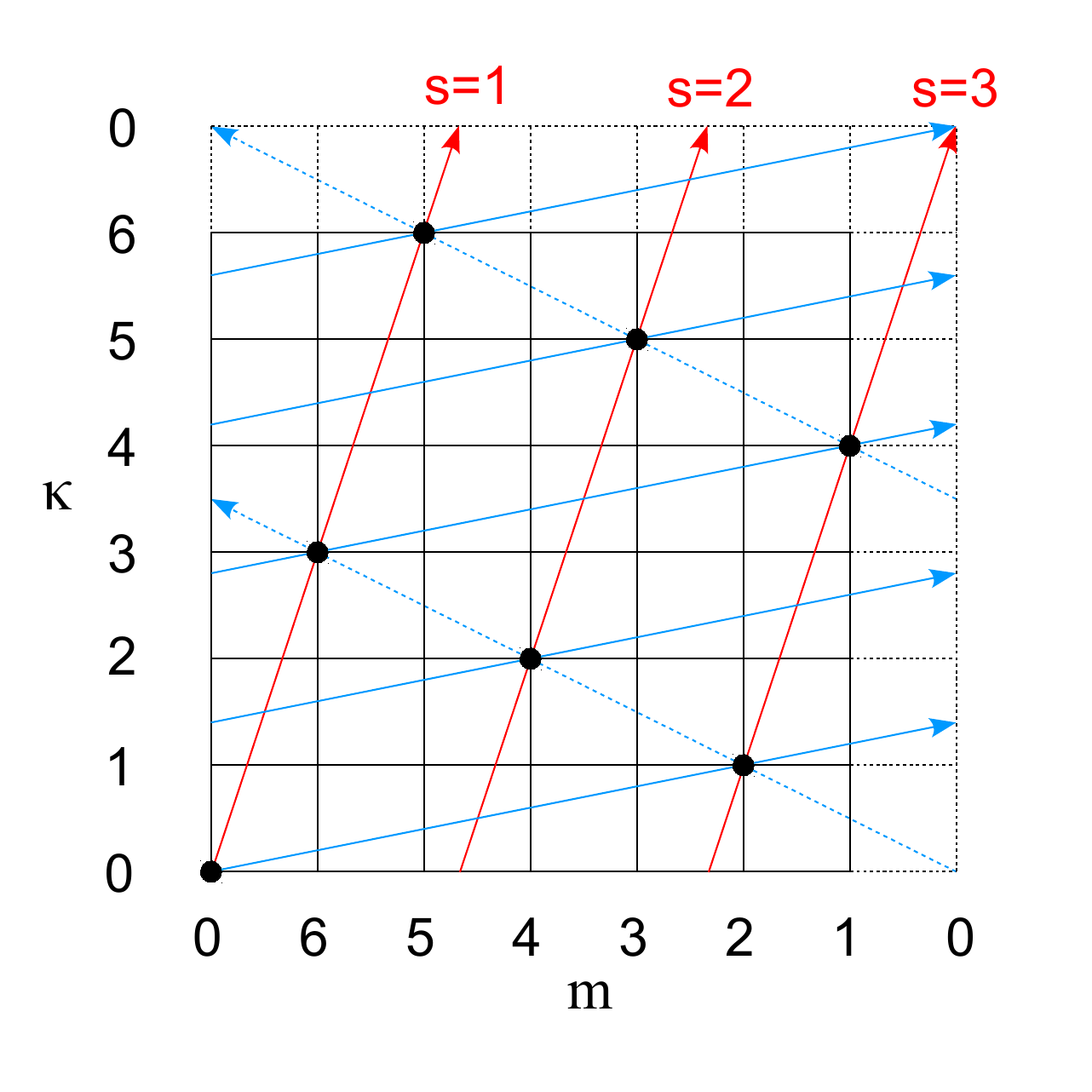}
\caption{(Color online) $(m,\kappa)$ construction for $p/q = 3/7$, assuming $m = 0$ when $\kappa = 0$.  Red line: flow in the
order of $T_y$ translation whereby $\kappa \rightarrow \kappa + p$ and $y \rightarrow y-1$. Blue lines: flow in the order of $k_x$
pumping. The number of time a packet is transferred across any boundary line, {\it e.g.\/}\ $m=0$ line, is the same as the number
of blue flow lines crossing the boundary line. There is a mod $q$ ambiguity as can be seen from the validity of both the solid and
dashed blue flows.}
\label{kappa-q}
\end{figure}

We now establish a connection between the band projectors and the
seminal work of Thouless {\it et al.\/}\cite{tknn82} on the Chern
numbers for the Hofstadter bands. Consider first the individual band
projectors $B_j$. We write
\begin{gather}
  y = q\ell+ m \quad , \quad k_x = {2\pi\kappa\over q}\ ,
\end{gather}
where $\ell$ and $m$ are integers. Thus $\ell$ is the magnetic unit
cell coordinate, and $m$ the coordinate within each such cell. For a
single band, denote the position of any of its wavepackets as $m(\kappa)$,
then $(k_x,y)$ covariance implies
\begin{gather}
  \label{kq}
  m(\kappa + tp) = m(\kappa) - t \quad, \quad t \in \mathbb{Z}\ .
\end{gather}
Of course both $m$ and $\kappa$ are only defined {\it modulo\/} $q$.
The relevant quantity in the $k_x$ pumping is the `velocity' of the
packet (with $k_x$ as `time'), {\it i.e.\/}, the number of sites it
traverses when $\kappa$ is \emph{effectively} increased by $1$:
\begin{gather}
  \label{diophantine-r1}
  tp = sq + 1 \quad , \quad s \in \mathbb{Z} \ , \ |t| < q\ .
\end{gather}
A graphical construction is shown in Fig.~\ref{kappa-q}. Clearly, $t$
will be the number of packets transported through any given boundary
during the cycle $k_x \rightarrow k_x + 2\pi$ (number of blue flow
lines in the figure). It is also equal to the number of magnetic cells traversed
by a single packet. There is however a mod $q$ ambiguity associated with
the sign indeterminancy of $t$, {\it e.g.\/} for $p/q = 3/7$, one has
that $(t,s) = (5,2)$ and $(-2,-1)$ both satisfy
eqn.~\ref{diophantine-r1}. In general, without looking at intermediate
$\kappa$ snapshots, one cannot tell if the packet had advanced by $t$ or
retreated by $q-t$. We have examined different $p/q$ ratios on both
square and triangular lattices, and we find that for the lowest band
on a square lattice, the ambiguity can always be resolved, without
needing to inspect intermediate $\kappa$, by picking the value of $t$
which has the smaller magnitude $|t|$, {\it e.g.\/} $t=-2$ instead of
$5$ for $B_1$ in Fig.~\ref{bigfig}. Intuitively, this means the packet
moves toward the nearest possible position allowed by eqn.~\ref{kq}.

Eqn.~\ref{diophantine-r1} is recognized as the Diophantine equation of TKNN\cite{tknn82} with $r=1$,
according to which $t$ is simply $C_1$, the Chern number of the lowest band.  The heuristic of taking the
smaller $|C_1|$ in resolving the mod $q$ ambiguity agrees with ref.~\onlinecite{tknn82}.  Since this
picture does \emph{not} distinguish between different bands, the Chern numbers of all
bands are equivalent mod $q$.

The correlation matrix $\CG_\nu$ for $\nu$ occupied bands has $\nu$
packets per diagonal block, with each moving according to
eqn.~\ref{diophantine-r1}, as required by the $(k_x,y)$ covariance.
However, their {\it collective\/} motion depends on their relative
spacing. Consider for example the diagonal blocks of $\CG_3$ for
$\kappa = 1$, as shown in top right panel of Fig.~\ref{bigfig}. In
each diagonal block, the diagonal elements with dominant weight
(packets) are at $y = (2,4,7)$. Then $(k_x,y)$ covariance requires
that at $\kappa = 2$, they are moved to $y = (2-t, 4-t, 7-t) = (4, 6,
9)$ where $t = -2$ is the solution to eqn.~\ref{diophantine-r1} as
discussed before. Since $y=9$ is simply the $y=2$ element of the next
diagonal block, thus in each diagonal block, the packets of $\kappa =
2$ are at $y = (4,6,9\ \mod\ 7 = 2)$. Comparing this with those of
$\kappa = 1$, one can see that, effectively, only one packet moved
from $y = 7$ to $y = 6$.

We found that for any $\nu \neq q$, this
observation holds true (for $\nu = q$, $\CG_{\nu}$ is identity). That
is to say, as $\kappa \rightarrow \kappa + 1$, the net effect is for
only {\it one\/} of the $\nu$ packets to change position.
We will not attempt to explain this observation,
but rather take it as a starting point, and explore
its implications. Mathematically, this observation -- the reduction of
the motion of multiple wavepackets to that of a single
mobile packet at a time -- means that the positions of dominant diagonal elements
(the packets) can be labeled in such a way that
\begin{gather}
\label{dot-recursion}
m_i(\kappa + 1) = m_{i+1}(\kappa) \quad , \quad i = 1, 2, \ldots, \nu-1
\end{gather}
with each $m_i$ still satisfying eqn.~\ref{kq}, {\it i.e.\/}
$m_i(\kappa + tp) = m_i(\kappa) - t$. Note that this does \emph{not}
mean $m_i$ can be identified with the wavepacket of a single band:
they result from constructive superposition of single-band projectors,
as discussed earlier.

To illustrate eqn.~\ref{dot-recursion}, take
again $\CG_3$ as an example: at $\kappa = 1$, $(m_1, m_2, m_3) =
(7,2,4)$, while at $\kappa = 2$, $(m_1, m_2, m_3) = (9 \ \mod \ 7 = 2,
4, 6)$. Then as $\kappa$ increases by $1$, the effective change is of
one packet (the mobile one) moving from $m_1(\kappa)$ to $m_{\nu}(\kappa + 1)$ with stride $t_{\nu}$,
\begin{gather}\label{stride}
m_\nu(\kappa + 1) = m_1(\kappa) - t_\nu \quad, \quad |t_\nu| < q\ .
\end{gather}
The RHS is therefore $m_1(\kappa + t_\nu p)$, while the LHS is
\begin{align}
m_\nu(\kappa + 1) = m_{\nu-1}(\kappa + 2) =\cdots = m_1(\kappa + \nu)
\end{align}
from eqn.~\ref{dot-recursion}.  Thus $t_\nu$ is determined by
\begin{gather}
\label{diophantine}
t_\nu p  = sq + \nu \quad , \quad s \in \mathbb{Z}\ ,\ |t_\nu| < q\ .
\end{gather}
Again, there is a mod $q$ ambiguity because of the sign indeterminancy
of $t_\nu$. For square lattice, the heuristic of using the smaller
$|t_\nu|$ still seems to hold, {\it e.g.\/} while both $(t_3, s) =
(1,0)$ and $(t_3,s) = (-6, -3)$ satisfy eqn.~\ref{diophantine}, the
actual system picks $t_3 = 1$. Ref.~\onlinecite{thouless84} mentioned
that $s$ and $t_\nu$ cannot simultaneously be odd for either the
hexagonal or triangular lattices. Incidentally, for $\nu=p$, $t_{p} =
1$ is always a solution with the corresponding $s = 0$, {\it i.e.\/}
the total Chern number of the lowest $p$ bands is always $1$.

Eqn.~\ref{diophantine} is the TKNN Diophantine equation\cite{tknn82}
for $r=\nu$. There, $t_\nu$ is the total Hall conductivity (the sum of
the Chern numbers) of the $\nu$ occupied bands. It is also the winding
number of the energy edge states in the $\nu^{\rm th}$
gap\cite{hatsugai93}. We now have a third interpretation: it is the
number of sites traversed by the mobile packet during each $\kappa$
increment. Equivalently, it is the number of mobile packets transported
across any magnetic unit cell boundary during the cycle $k_x
\rightarrow k_x + 2\pi$.

The entanglement spectrum can now be understood intuitively.  Whenever the mobile packet leaves the lower
half-cylinder through the cut between $M$ and $M+1$ (in Fig.~\ref{bigfig}, proceeding from top-left quadrant
through the red line into the lower-right quadrant), there is an occupancy flow from $f=1$ to $0$.
The number of flow lines is then equal to the number of packets which move through the cut,
which is the total Chern number. In the periodic $y$ boundary case, where the cylinder is compactified into a torus,
the flow across $M = N_y/2$, is always concomitant with another packet moving from $y=N_y$ to $y=1$, hence a symmetric
flow from $0$ to $1$ with its wavefunction localized at the opposite end.  (Entanglement occupancy with periodic $y$
boundary is shown in Fig.~\ref{wc-g} in the next section).  Furthermore, If we change the position of the entanglement
cut $M$ (not necessarily along a magnetic cell boundary, for example), this will simply change
the value of $k_x$ when a packet hits the cut, whence the $k_x$ translation shown in Figs.~\ref{p3q7ny28}
and \ref{p3q7ny29}.

While the entanglement spectrum only reveals the total Chern number,
the correlation matrix retains some information about the individual
Chern numbers of constituent bands, manifested as the separation
between its wavepackets. Note that eqn.~\ref{stride} can be taken as a
definition of $t_\nu$ with arbitrary $\nu < q$, without interpreting
$\nu$ as the number of filled bands.  After all, the total Chern
number of the lowest $\nu$ bands is the same whether or not they are
filled.  We explicitly replace $\nu$ with $n < q$ below to avoid any
such connotation. From eqn.~\ref{dot-recursion} and \ref{stride}, we have
\begin{gather}
  \label{dot-distance}
  m_{n+1}(\kappa) - m_n(\kappa) = t_{n-1} - t_n = - C_n\ ,
\end{gather}
thus the two packets at $m_{n+1}$ and $m_n$ are separated by a distance
of $C_n$. Quantities such as the four-point correlation
$\mathcal{F}(\Delta) = \langle \, c\yd_y \, c\nd_y \, c\yd_{y+\Delta}
\, c\nd_{y+\Delta} \, \rangle$ thus have peaks at $\Delta = C_i$ apart
from $\Delta = q$, $2q$, {\it etc.\/}

On a square lattice, the Hofstadter model exhibits a particle-hole
symmetry. This implies that $C_j = C_{q+1-j}$. For even $q$, the bulk
spectrum is known to have no central gap\cite{wen89-zm,komoto89-zm},
therefore the Chern numbers of the two central bands are not
individually well defined, and one can speak only of a Chern number
for the pair. It is interesting to notice its implication on the
distribution of the wavepackets within each unit cell: if on the
contrary there is a central gap, then $t_{q/2} = C_1 + C_2 +
\ldots + C_{q/2} = C_{q} + C_{q-1} + \cdots C_{q/2+1}$.
Since the total Chern number of all bands must be zero, we must have
$t_{q/2} = 0$. Now according to eqn.~\ref{dot-recursion} and
\ref{stride}, $m_{{q/2}+1}(\kappa) = m_1(\kappa) - t_{q/2}=
m_1(\kappa)$, so the $({q\over 2}+1)^{\rm th}$ packet and the first
one are forced onto the same site. Thus, the fact that there is {\it
  no\/} central gap in this case guarantees that there will be no
packet `collisions'.

The natural question to ask next is how the wave
packets are arranged when $\nu > q/2$. To illustrate this, consider a
specific case with $p/q = 5/8$ and $\nu = 7$ filled bands. The Chern
numbers of the lowest three bands are $-3$, $5$, and $-3$,
respectively, and particle-hole symmetry guarantees that these values
repeat for the upper three bands. The central two bands therefore have
a combined Chern number of $C_{4,5} = 2$. Since Chern numbers
represent the separation between wave packets, we can fill in the
first four packets with no difficulty (the position of the first
packet being arbitrary). The location of the fifth packet cannot be
determined because $C_4$ is not well-defined, but the location of the
sixth packet is found by shifting the fourth one by $C_{4,5} = 2$. The
rest of the packets can be filled in a similar fashion. Thus the
vanishing of the central gap implies an indeterminacy of the position
of the $({q \over 2} + 1)^{\rm{th}}$ wavepacket. To resolve this, one
can add in an infinitesimal second-neighbor hopping that breaks the
particle-hole symmetry and results in a small central gap. For
example, one can introduce a second-neighbor hopping $t'$ along one of
the two diagonals in each unit cell (say in the direction ${\hat x} -
{\hat y}$). This construction interpolates between the square lattice
when $t'=0$, and the triangular lattice when $t'=1$ (see
Appendix~\ref{triangle}). For $t'\ll 1$, we find that the Chern
numbers of the central two bands are resolved as $C_4 = 5$ and $C_5 =
-3$. $C_4$ can now be used to determine the position of the
fifth packet.

\section{Wannier center flow}

The Diophantine equation (\ref{diophantine}) describes a mod-$q$ property of the Hofstadter problem,
which is a result of the $(k_x,y)$ covariance of eqn.~\ref{covariance}.  No knowledge of intermediate
values of $k_x\in [\,\kappa \phi\, , \, (\kappa + 1) \phi\,]$ is required in obtaining
eqn.~\ref{diophantine}. This comes at a price of the ambiguity in $t$ (mod $q$) and $s$ (mod $p$),
which intuitively contain the information of the direction in which any given packet is moving.
In this section, we settle this issue by examining the
localized eigenstates of the projectors over the full range of $k_x$, {\it i.e.\/} the Wannier functions.

\subsection{Wannier functions in 1D}
\label{wf1d}
The application of Wannier functions to the analysis of topological band structures has recently been developed in refs.
\onlinecite{soluyanov-vanderbilt11-z2-wannier}, \onlinecite{Yu11-z2-wannier}, and \onlinecite{Qi11-wannier}.
Following these references, consider first a periodic one-dimensional system consisting of $N$ unit cells with $q$
internal degrees of freedom per cell.  Let $X$ be a cell coordinate and let $m$ index the internal degree of freedom;
we may take $X\in\{1,\ldots,N\}$ and $m\in\{1,\ldots,q\}$.  Bloch's theorem says
$\Psi_{n,k}(X,m)=e^{i kX}\,u_{n,k}(m)$, where $n$ labels the $q$ bands. One may thus decompose the Hilbert space as
${\cal H}={\cal H}_X\otimes{\cal H}_m$, writing $|\Psi_{n,k}\rangle = |k\rangle\otimes |u_{n,k}\rangle$.  Here
$|u_{n,k}\rangle$ is an eigenstate of the Fourier transform $H_{mm'}(k)$ of
$H_{mm'}(X-X')\equiv \langle X,m\, | {\hat H} | \, X',m'\rangle$, {\it i.e.\/} it is a Bloch cell function.

For a system with periodic boundary conditions, the position operator can be taken to be $U=e^{2\pi i{\hat X}/N}$, as in
the work of Yu {\it et al.\/}\cite{Yu11-z2-wannier}  An eigenstate of ${\widetilde U}\equiv P\, U P$, where $P$
is a projector onto a subset of energy bands, is of the form 
\begin{gather}
|\Phi_\lambda\rangle =\sum_{n,k} \Phi^\lambda_{n,k}\,|k\rangle \otimes |u_{n,k}\rangle\ ,
\end{gather}
where the sum on the band index $n$ is over the desired subset, and $\lambda$ labels the eigenvalues.  
Demanding
${\widetilde U}\, |\Phi_\lambda\rangle = e^{2\pi i\lambda/N}|\Phi_\lambda\rangle$,  one obtains
$\Phi^\lambda_{m,k+\Delta k}=e^{-2\pi i\lambda/N}\,M_{mn}(k+{1\over 2}\Delta k)\,\Phi^\lambda_{n,k}$ (sum on $n$ over selected bands), where
\begin{gather}
M_{mn}(k)=\langle \, u_{m,k+{1\over 2}\Delta k} \, | \, u_{n,k-{1\over 2}\Delta k} \, \rangle\ ,
\end{gather}
with $\Delta k=2\pi/N$.  The eigenvalue equation, which follows from setting $\Phi^\lambda_{n,0}=\Phi^\lambda_{n,2\pi}$, is then
\begin{gather}
\textsf{det}\big( e^{2\pi i\lambda} - W \big) = 0\ ,
\end{gather}
where $W=M(N\Delta k - {1\over2}\Delta k)\cdots M(\Delta k - {1\over
  2}\Delta k)$ is a Wilson loop.  
Note that $\lambda$ is not necessarily real since
${\widetilde U}$ is the {\it projection\/} of a unitary operator but is not unitary itself.  In a more general setting,
where the wavefunctions $u_m$ depend on a set of parameters ${\vec g}$, one has
\begin{align}
&\big\langle u_m({\vec g}+\half\Delta{\vec g}) \, \big| \, u_n({\vec g}-\half\Delta{\vec g}) \big\rangle \\
&\qquad
=\Big[ \exp\big(i A^\mu \Delta g_\mu - \frac{1}{2} Q^{\mu\nu} \Delta g_\mu \, \Delta g_\nu + {\cal O}(\Delta g ^3)\big)\Big]_{mn}\ ,
\nonumber
\end{align}
where $A^\mu$ is the nonabelian Berry connection,
\begin{gather}
A^\mu_{mn}({\vec g})=i\,\Big\langle u_m \, \Big| \, {\partial u_n\over\partial g_\mu}\Big\rangle
\end{gather}
and $Q^{\mu\nu}$ is the quantum geometric tensor\cite{provost80,zanardi07},
\begin{gather}
Q^{\mu\nu}_{mn}({\vec g})=\Big\langle {\partial u_m\over\partial g_\mu} \, \Big|\, (1-P) \, \Big|
\, {\partial u_n\over\partial g_\nu}\Big\rangle \ .
\end{gather}
In our case, as $N\to\infty$ we have that the Wilson loop becomes unitary, and each eigenvalue $\lambda$ is real.

For a single band, we can write
\begin{gather}
  \lambda_I = \int\limits_0^{2\pi} \! {dk\over 2\pi} \,A(k) + I\ ,
\end{gather}
where $I$ is an integer and $A(k)=i\,\langle \, u(k) \, | \,{\partial\over\partial k} \, | \, u(k) \,\rangle$.
Thus for a single band, the state $|\Phi_{\lambda_I}\rangle$ is localized at unit cell $I$ with an offset
$\gamma/2\pi=\int\limits_0^{2\pi}\! dk\, A(k)$.

When the internal space of $|u_{n,k}\rangle$ coincides with real
space ({\it e.g.\/} the lattice site $m$ within the magnetic unit cell in
Hofstadter problem), one may refine the definition of the position operator, writing
\begin{equation}
U\to e^{2\pi i {\hat X}/N}\,e^{2\pi i {\hat m}/qN}\ ,
\label{Uref}
\end{equation}
where ${\hat m}=\sum_{m=1}^q m\,|m\rangle\langle m |$ measures the position within each unit cell.
For the single band case, this shifts the offsets $\gamma$ to
\begin{equation}
{\tilde\gamma}=\gamma + q^{-1} \! \int\limits_0^{2\pi} \! dk ~\langle u_{n,k} | \, \hat m \, |u_{n,k}\rangle\ .
\end{equation}
Equivalently, one may also introduce the modified cell functions
$|{\tilde u}_{n,k}\rangle$,
\begin{gather}
|{\tilde u}_{n,k}\rangle = q^{-1/2}\sum_{m=1}^q \langle m \, | \, u_{n,k}\rangle \,e^{-ikm/q}  \, |m\rangle \, 
\end{gather}
and use them in computing the Berry connection.

When the projector $P$ is onto multiple bands, the Wilson loop becomes a matrix, and the eigenvalues of ${\widetilde U}$
are $e^{2\pi i\lambda_{I,w}/N}$, where $w$ is an additional label running from one to the number of bands, {\it i.e.\/}
the dimension of the projector\cite{Qi11-wannier}.  One then has
\begin{equation}
\lambda_{w,I}=I+{\theta_w\over 2\pi}\ .
\label{lambda-wi}
\end{equation}
Again, for systems where the internal `orbital space' corresponds to real space, one can refine the position operator as
in eqn. \ref{Uref}.

For two-dimensional lattices, Wannier functions can be defined at each $k_x$. For a
single band $n$, its $y$-center is the Berry phase $\gamma_n(k_x)/2\pi$.
The band Chern number is the negative of the winding number of
$\gamma_n(k_x)$ over the interval $k_x\in[0,2\pi]$ :
\begin{align}
  C_n &= \frac{1}{2\pi i} \int\limits_{\rm BZ} \! d^2\!k\> {\vec\nabla}_{\!\vec k} \times \langle
  \psi_n(\vec k) | \, {\vec\nabla}_{\!\vec k} \, | \psi_n(\vec k)\rangle \cdot{\hat z} \nonumber\\
  &= {\gamma_n(0) - \gamma_n(2\pi) \over 2\pi}\ .
\end{align}
Thus the Wannier center shifts by $-C_n$ (magnetic) unit cells over
$k_x \rightarrow k_x + 2\pi$, as found by Qi in ref. \onlinecite{Qi11-wannier}.
We have seen in \S\ref{sec-proj} that the packet associated with band $n$ is translated by $-C_n$
lattice sites during $k_x \rightarrow k_x + 2\pi/q$, so over $k_x
\rightarrow k_x + 2\pi$, it will be translated by $-C_n$ magnetic unit
cells, in agreement with the Wannier picture.

For multiple bands, we have 
\begin{gather}
\textsf{det}\, W = \exp\left\{i \!\int\limits_0^{2\pi} \!dk \> \textsf{Tr} \> A(k) \right\}
\end{gather}
hence
\begin{gather}
  \label{sum}
  \sum_{w=1}^{\nu} \theta_w = \sum_{n = 1}^{\nu} \gamma_n\ ,
\end{gather}
where $\nu=\textsf{rank}(P)$.
Then similar to the single band case, we conclude that in $2$D, the
total shift of all (inequivalent) Wannier centers is given by the sum
of the Chern numbers.  This is reflected in \S\ref{sec-proj} as the number
of mobile packets transported through any given magnetic unit cell boundary.

When there is no level crossing among $\{\lambda_{w,I}\}$ over the
period of $k_x$, one can combine the $w$ and $I$ indices.  Define a composite index
$\mu(w, I) = w+ \nu I $, with $\lambda_{w,I} \rightarrow \lambda_{\mu}$.
Sending $k_x$ from $0$ to $2\pi$ amounts to an index shift, which
is universal for all $\theta_{\mu}$ since there is no level crossing.  Then the
eigenfunctions $\Theta_\mu(k_x)$ of $W$ (here taken to be periodic in the index {\it modulo\/} $q$) satisfy
\begin{gather}
  \Theta_{\mu}(2\pi) = \Theta_{\mu + \sigma}(0)  \quad, \quad
  \sigma \in \mathbb{Z}
\end{gather}
which is just a cyclic permutation in the $w$ index with an offset
$\sigma$.  One may think of the set of $\{e^{i\theta_w}\}$ as $\nu$
points on the unit circle where different indices $I$ are equivalent.
Then during the $\sigma$ cyclic permutation, the perimeter of the
circle is covered by these $\nu$ points for $\sigma$ times, {\it i.e.\/},
\begin{gather}
{1\over 2\pi}\sum_{w=1}^\nu \big[\theta_{w,I}(2\pi) - \theta_{w,I}(0)\big] = \sigma\ .
\end{gather}
But according to eqn.~\ref{sum}, the LHS is simply the total Chern
number of constituent bands,
\begin{gather}
  \label{mb-winding}
\sigma = -\sum_{n=1}^{\nu} C_n\ .
\end{gather}

\begin{figure*}[!t]
  \subfigure[$\quad 1$ filled bands]{
    \label{wc-b1}\includegraphics[width=0.35\textwidth]{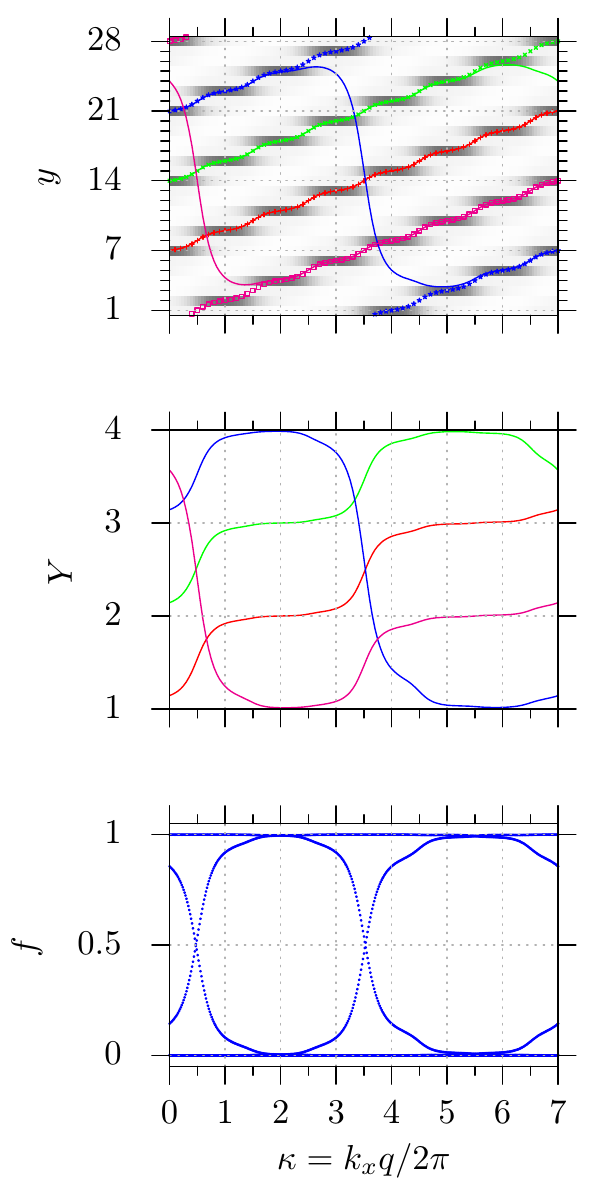}
  }\subfigure[$\quad 2$ filled bands]{
    \label{wc-g2}\includegraphics[width=0.28\textwidth]{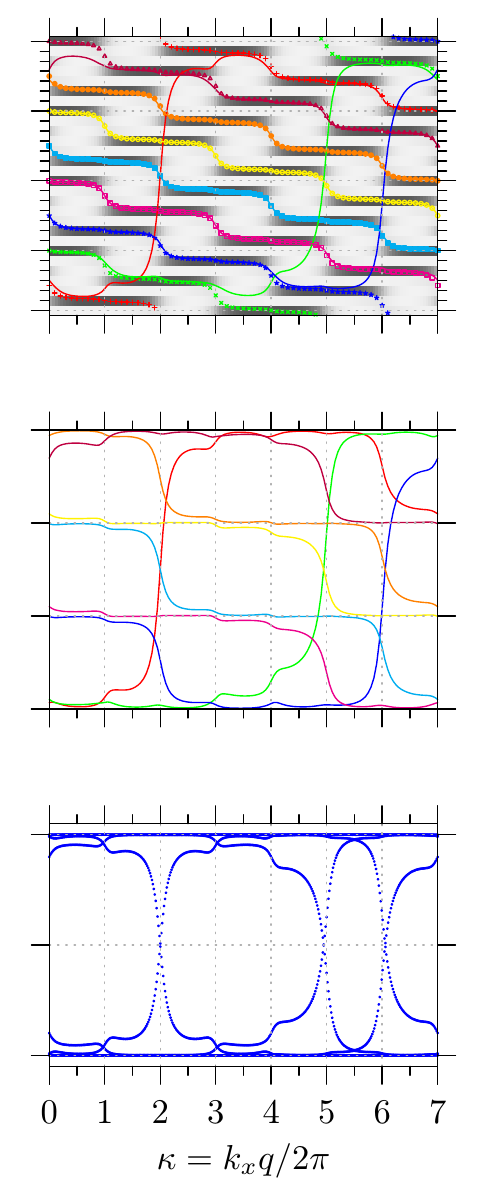}
  }\subfigure[$\quad 3$ filled bands]{
    \label{wc-g3}\includegraphics[width=0.28\textwidth]{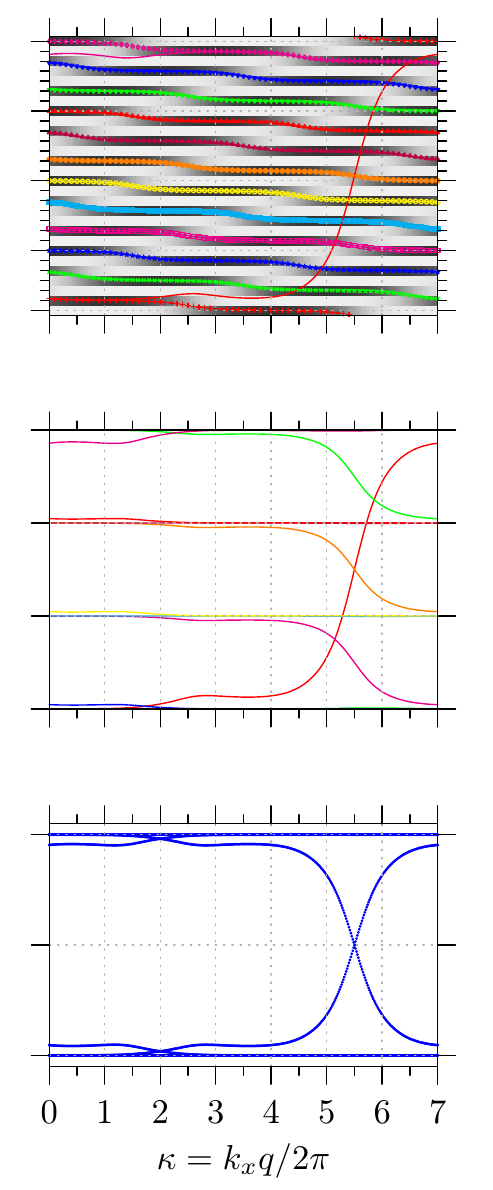}
  }\caption{(Color online) Wannier centers {\it vs.\/} entanglement
    occupancy for $p/q = 3/7$ with different filling fraction and $4$
    magnetic unit cells in the full system and $2$ in the half system.
    Top: Wannier centers using lattice coordinate $\hat y$ as position
    operator. Colored dots: $y_\theta$. Colored lines: $\langle
    y\rangle_\theta$. Black-white background: diagonal elements of the
    lowest band projector $B_1$, black $= 1$, white $= 0$. See also
    Fig.~\ref{bigfig}. Center: Wannier center using $\langle Y
    \rangle_\theta$ (magnetic unit cell coordinate). Corresponding
    levels have the same color as in top panels. Bottom: entanglement
    occupancy (no type/color coding). In (a), each type/color of point
    in the top and center panels corresponds to a packet, \emph{i.e.},
    related via $(k_x,y)$ covariance, in this case advancing by $p=3$
    in $\kappa$ as $y \rightarrow y+6$ (equivalent to $y-1$). In (b)
    and (c), each type/color of point corresponds to a mobile packet
    (because they are not related by the $(k_x,y)$ covariance).
    Specifically, in (b), advancing $\kappa$ by $p=3$ does not change
    $y$ by $-1$ or $6$, and in (c), advancing $\kappa$ by $1$ does not
    change $y$ by $2$ or $-5$. It is also clear that within each
    magnetic cell, only one mobile packet moves as $\kappa \rightarrow
    \kappa + 1$, with $\kappa = $ half-odd-integer in (b) and integer
    in (c). Notice the similarity between the Wannier center flow and
    the entanglement flow: a packet (a) or mobile packet (b and c)
    crossing the magnetic cell within the bulk has almost the same
    shape as the entanglement downflow; while crossing from $y=N_y$ to
    $y=1$ has the almost the same shape as the entanglement upflow.
    The plateau-like feature in the entanglement flow lines can be
    traced back to the $y$ plot in the top panels as the motion of a
    (mobile) packet within one magnetic cell.}
  \label{wc-g}
\end{figure*}

\begin{figure}[!t]
\begin{center}
\includegraphics[width=9cm,angle=0]{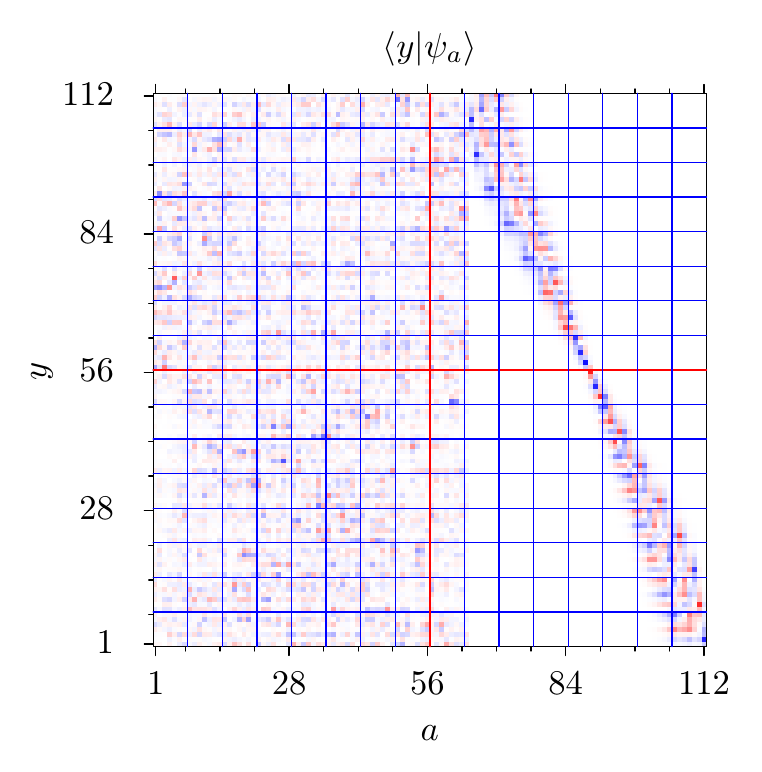}
  \caption{(Color online) Normalized eigenfunctions $\psi_a(y)$ of $\CG R\CG$ for $p/q=3/7$, $N_y=112$,
  $M=56$, and $\nu=3$ at $k_x=2\pi/7$ (cylindrical boundary conditions).  The Chern number of the filled bands is $\Cocc=1$.
  Color corresponds to the sign of the wavefunction (red for positive, blue for negative), and intensity
  to amplitude (white for zero intensity).}
\label{PRPwfs} 
\end{center}
\end{figure}

\subsection{Wannier center flow in Hofstadter problem and a general relation with the entanglement spectrum}
For the Hofstadter model, we have numerically diagonalized the operator ${\widetilde U}=P\, e^{2\pi i {\hat Y}/N_y}\,
e^{2\pi i {\hat m}/qN_y} P$, where $Y\in\{1,\ldots,N_y\}$ runs over the unit cells, and $m\in\{1,\ldots,q\}$ runs over the
individual sites within each unit cell. States with zero eigenvalues are those projected out by $P$.  For a single band, $P=B_n$,
while $P=\CG_\nu$ for $\nu$ filled bands. We write ${\widetilde U}\,|\theta\rangle = e^{i\theta}\,|\,\theta\,\rangle$,
and we compute three slightly different Wannier centers: $y\nd_\theta\equiv\theta N_y/2\pi$, $\langle Y\rangle\nd_\theta\equiv
\langle\,\theta\,|\,{\hat Y} \, |\, \theta\,\rangle$, and $\langle y\rangle\nd_\theta\equiv
\langle\,\theta\,|\,{\hat Y} + {{\hat m}\over q} \, |\, \theta\,\rangle$, where $\theta$ is wrapped in such a way that
$y\nd_\theta$ is restricted to $[0.5, N_y + 0.5]$. The top panel of fig.~\ref{wc-b1} shows the Wannier
centers defined above for the lowest band of $p/q=3/7$ with $N_y = 28$.  The $y\nd_\theta$ values, shown as colored dots,
have a proper translational property: these values are generated by shifting any single flow by successive multiples of $q$.
In the vicinity of half-odd-integer $\kappa$, a Wannier center migrates from one site to the next one which is $-C_1 = 2$
sites ahead. In terms of the corresponding wavefunction (not plotted), what happens is that around integer $\kappa$,
it has a single peak at the site given by its (rounded) eigenvalue $y\nd_\theta$.  As $\kappa$ slowly moves toward the next
half-odd-integer value, some weight is transferred to the next site, causing the eigenvalue $y\nd_\theta$ to interpolate between
the two values. As $k_x$ is increased by $2\pi$, $q$ such migrations are made, {\it i.e.\/}, each Wannier center is shifted
backwards by $C_1$ magnetic unit cells. In the bulk, $\langle y \rangle\nd_\theta$ (colored lines in (a)) and $y\nd_\theta$ overlap.
Near $y=N_y$, part of the weight is pushed over the boundary to the $y=1$ end, thus $\langle y\rangle\nd_\theta$
starts to deviate from $y\nd_\theta$ and drops, until all weight is transferred to the other boundary.

If instead of ${\widetilde U}$, we were to diagonalize the operator
${\widetilde V}\equiv P\left({{\hat Y}\over N_y} + {{\hat m}\over q N_y}\right)P$,
then the behavior in the bulk will be the same, but near the edge there will be avoided crossings
in the Wannier center flow of $\langle y\rangle\nd_\theta$. The flow of the magnetic unit cell coordinate
$\langle Y \rangle\nd_\theta$, shown as colored lines in fig.~\ref{wc-g2}, is similar to that of
$\langle y\rangle\nd_\theta$, but with an emphasis on the occasions
when $\langle y\rangle\nd_\theta$ crosses a magnetic cell boundary.

\begin{figure}[!t]
\begin{center}
\includegraphics[width=8.7cm,angle=0]{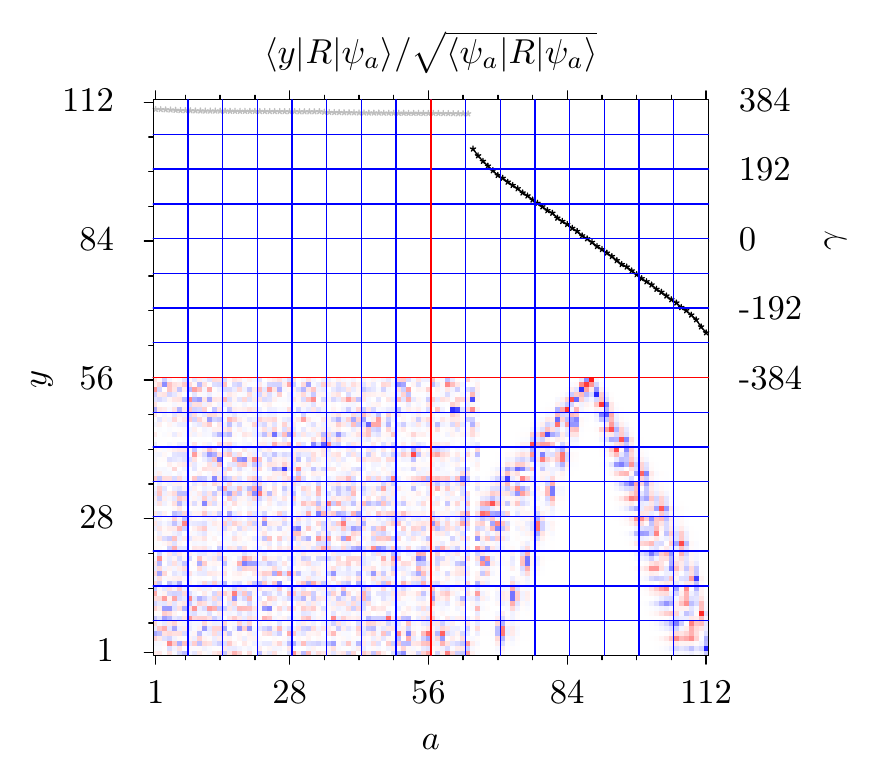}
  \caption{(Color online) Normalized eigenfunctions ${\tilde\psi}_a(y)$ of $R\CG R$ for the parameters given in
  the caption to Fig. \ref{PRPwfs}.  The entanglement energies are plotted in the top half.  Grey points correspond
  to $\gamma_a=\infty$ ($f_a=0$).  As expected, states with entanglement energy $\gamma_a\approx 0$ are localized
  in the vicinity of the cut.}
\label{RPRwfs} 
\end{center}
\end{figure}

The Wannier centers of two and three filled bands are shown in
Fig.~\ref{wc-g}. As $\kappa$ increases by $1$, only one Wannier center
per magnetic unit cell flows by $\sigma=-\sum_{i=1}^\nu C_i$ sites (Eq. \ref{mb-winding}), corresponding to the
motion of a mobile packet in \S\ref{sec-proj}. $\sigma=-3$ for two filled
bands and $1$ for three filled bands. There is no level crossing in
$y_{\theta}$, hence eqn.~\ref{mb-winding} holds true: following any
flow line from $k_x = 0 $ to $2\pi$ leads one to $-\sigma$ levels
beneath the starting point. Notice the same behavior in the
entanglement quasienergy of Fig.~\ref{c2-gamma}.

There is a striking similarity between the magnetic cell coordinate flow
$\langle Y \rangle$ and the entanglement occupancy spectrum, shown in
the center and bottom rows of Fig.~\ref{wc-g}: the upward flow of
$\langle Y \rangle$ looks exactly like the upward flow of $f$, while
the downward flow of $\langle Y \rangle$ within each $Y \rightarrow
Y-1$ sector looks like the downward flow of $f$. This can be
understood in the following way: the spectrum of $\hat Y$ can be
thought of as a coarse-grained version of $\hat y$, and should look
just like $\langle Y \rangle$ and $\langle y \rangle$ shown in
Fig.~\ref{wc-g}, so the effect of coarse-graining is to suppress the
flow within a coarse-grained cell, and enhance the flow migrating
between different cells. One may push the coarse graining to the
extreme where all sites with $y \le M$ count as $\bar y=0$, and all
above $M$ as $\bar y=1$. Then the only significant flow is from
$\bar y = 0$ to $1$, corresponding to $Y=2$ to $Y=3$ in
Fig.~\ref{wc-g}, and from $\bar y = 1$ to $\bar y = 0$, corresponding to $Y=4$
through the periodic boundary to $Y = 1$. In Fig.~\ref{wc-g}, such a
coarse graining would keep the upward $Y$ flow intact, while push
$Y=3$ and $Y=2$ lines to top and bottom respectively for the downward
flow, making it look just like the entanglement occupancy spectrum. In
fact, one can prove that the coarse-grained Wannier spectrum is
\emph{identical} to the entanglement spectrum: Consider two arbitrary
projectors $P$ and $R$. One can think of them as two matrices of the
same dimension (zero-padded, if necessary, to fill out the dimensions). We claim that $PRP$ and
$RPR$ have identical eigenspectra. To see this, assume $| \psi_a \rangle$
is an eigenstate of $PRP$ with non-zero eigenvalue $\lambda_a$. Then
\begin{gather}
  |\psi_a\rangle = \frac{1}{\lambda_a} \,PRP\, |\psi_a\rangle
\end{gather}
and therefore $P |\psi_a\rangle = |\psi_a\rangle$.  Thus, $|\psi_a\rangle$
is an eigenstate of $P$ with eigenvalue $1$.  It then follows that 
\begin{gather}
 RPR\, |\psi_a\rangle = \lambda_a R\, |\psi_a\rangle\ ,
\end{gather}
from which it follows (using $R^2=R$) that $|{\tilde\psi}_a\rangle=R\,|\psi_a\rangle$ is an eigenstate of $RPR$ with 
the same eigenvalue $\lambda_a$.  Thus the non-zero spectrum 
of $PRP$ belongs in that of $RPR$, and vice versa, so they are identical.  Since the coarse-grained position
operator $R$ is (the complement of) the projector used in constructing the restricted correlation matrix $G$, the entanglement
spectrum of $G= R \CG R$ is identical to the coarse-grained Wannier centers $\CG R \CG$, and the entanglement eigenstates are
obtained by projecting the coarse-grained Wannier states onto the relevant half space.

In Fig. \ref{PRPwfs} we plot the normalized eigenfunctions $\psi_a(y)$ of $\CG R\CG$ for the case $p/q=3/7$, $N_y=112$, $M=56$, and
$\nu=3$ at $k_x=2\pi/7$.  Note how the behavior of the eigenfunctions mimics that of the Wannier states, with $\psi_a(y)$ localized at
a position which moves across the entire cylinder as the label $a$ advances through a range corresponding to the rank of $\CG R \CG$
 (${3\over 7}\times 112 = 48$ in this case; see \S \ref{ranksec}).  States \#1 through \#64 belong to the kernel of $\CG R \CG$ and are
 all degenerate.  The insertion of the real space projector $R$ thus fails to resolve these wavefunctions in real space, which explains
 the speckled pattern on the left side of the figure.  The Wannier states are better localized however, since $R$ may be considered
 a  coarse-grained approximation to $y$.

In Fig. \ref{RPRwfs}, we plot the normalized eigenfunctions ${\tilde\psi}_a(y)$ of $R\CG R$ for the same parameters,
along with the corresponding entanglement energies $\gamma_a$. (If we remove the upper half of the cylinder, where the wavefunctions
vanish, this is a repeat of Fig. \ref{Gwfs}.)  According to our definitions,
\begin{gather}
|{\tilde\psi}_a\rangle = R\,|\psi_a\rangle \Big/ \sqrt{\langle \psi_a | \, R \, | \psi_a \rangle}\ .
\end{gather}
Note how states of large positive $\gamma_a$ ($f_a\approx 0$) as well as states of large negative $\gamma_a$ ($f_a\approx 1$) are localized
far from the $A/B$ boundary.  

\section{Summary}

We have studied the entanglement spectrum and Wannier center flows of the Hofstadter problem.
Most of the data presented in this paper was for the square lattice with $p/q=3/7$ flux quanta per unit cell,
but most of our observations are robust with respect to changing lattices, fluxes, and fillings.
The entanglement spectrum of a subsystem exhibits spectral
flow similar to that of the full system's energy edge modes: the total Chern number
controls the number of flow lines, and its sign tells the direction of
the flow.  When cylindrical boundary conditions are used in the full system, the
entanglement spectrum exhibits level index discontinuity on the flow
line.  This is a manifestation of the crossing of the Fermi energy with
the full system edge modes, which results in a total occupancy
discontinuity.  Changing the location of the entanglement cut shifts
the entanglement spectrum. This reflects the $k_x\;y$ covariance of
the Hamiltonian: changing $k_x$ to $k_x + 2\pi p/q$ is equivalent to
shifting the system in $y$ by $\Delta y=1$.

The behavior of the entanglement spectrum can be understood by looking
at the full system band projectors. These projectors are well
localized and thus represented by wavepackets on their diagonals.
For single bands, the packets flow under $k_x$ pumping.  The $(k_x,y)$
covariance then imposes restrictions on possible flow rate, described
by a Diophantine equation first derived by TKNN\cite{tknn82}.  Since the Chern numbers are also given by
the same equation, the topological property of the system can be
described equivalently in terms of the motion of these packets: the number of
magnetic unit cells traversed by each packet during one period of $k_x$
is given by the Chern number, with the direction given by its sign.
For multiple bands, the flow is that of the mobile packets moving under $k_x$
pumping, and the number of mobile packets crossing a given boundary gives
the total Chern number of filled bands.  The entanglement spectrum can
then be understood as a measure of detecting when these packets cross a
particular boundary, namely the entanglement cut.

Using the $(k_x,y)$ covariance alone (and hence the Diophantine
equations) only fixes the flow and the Chern numbers up to mod $q$
because it only relates different $k_x$ points of fixed separation of
$2\pi / q$. The localization of the projectors suggests the use of
Wannier functions for smooth interpolation between these $k_x$ points.
For single bands, the Wannier center at each $k_x$ is given by the
corresponding Berry phase and is represented by one packet in the
projector diagonal. The flow of the Wannier center is then described
by the winding number of this Berry phase, which is the band Chern
number. For multiple bands, the Berry phase is replaced by a set of
eigenvalues of the Wilson loop operator. If there is no level crossing
over the full range of $k_x$, then all levels experience a universal
index bump of $\sigma$ as $k_x \rightarrow k_x + 2\pi$, and $\sigma$
is given by the sum of Chern numbers. In computing the Wannier center,
the position operator can be either the (magnetic) unit cell
coordinate alone, or one that also includes the internal coordinates
(lattice cell within each magnetic cell). The spectrum of the former is a
coarse-grained version of the latter. One can take the coarse graining
to the extreme of a bipartition, at which point the position operator
becomes a real space projector, and the coarse-grained Wannier
spectrum becomes identical to the entanglement occupancy spectrum.

\section{Acknowledgments}
This work was supported by the NSF through grant DMR-1007028.
We thank B. A. Bernevig and A. Alexandradinata for useful discussions, and for comments after
reading a draft of this paper.

\appendix

\section{Incommensurate edge spectrum}
\label{incommensurate}
Here we first briefly review the edge spectrum with commensurate $N_y$
and $q$ as studied in ref.~\onlinecite{hatsugai93}, and then extend
its argument to the incommensurate case.

\subsection{Review of commensurate edge spectrum}
The Schr\"odinger equation corresponding to the matrix equation
$H(k_x)_{yy'}\psi_{y'} ~= \varepsilon \psi_y$ is
\begin{gather}
  -\psi_{y-1} - \psi_{y+1} - 2\cos(k_x + y\phi) \, \psi_y = \varepsilon
  \psi_y
\end{gather}
cast into transfer matrix form, we have
\begin{gather}
  \label{transfer}
  \begin{pmatrix}
    \psi_{y+1} \\ \psi_y
  \end{pmatrix} = M_y
  \begin{pmatrix}
    \psi_y\\\psi_{y-1}
  \end{pmatrix}, \\
  \label{tmat}
  M_y = \begin{pmatrix}
    -\varepsilon - 2\cos(k_x + y\phi) & -1\\
    1 & 0
  \end{pmatrix}
\end{gather}
Notice that $M_y$ depends on $\varepsilon$. The following boundary
condition is required for eqn.~\ref{transfer} to also cover the
cases of $y=1$ and $N_y$,
\begin{align}
  \label{bc}
  \psi\nd_0 = \psi\nd_{N_y+1} =0\ .
\end{align}
Then

\begin{gather}
  \begin{pmatrix}
    \psi_{N_y+2} \\ \psi_{N_y+1}
  \end{pmatrix} = \mm{N_y+1}
  \begin{pmatrix}
    \psi_1 \\ \psi_0
  \end{pmatrix}\\
  \mm{y} \equiv M_{y} M_{y-1} \cdots M_1
\end{gather}
and eqn.~\ref{bc} implies $M_{N_y+1}$ is a triangular matrix,
\begin{align}
  \label{constraint}
  \mme{N_y+1}_{21} = 0
\end{align}
The spectrum $\{\varepsilon\}$ consists of all energies satisfying
eqn.~\ref{constraint}.

Notice that $M_{y+q} = M_y$, so
when $N_y+1 = qL$ with integer $L$ (``commensurate''),
\begin{gather}
  \mm{N_y + 1} = \qq^{L}\quad , \quad \qq \equiv \mm{q}
\end{gather}
Now, products of up-triangular matrices are still up-triangular, so
eqn.~\ref{constraint} is satisfied if $\qq$ is up-triangular,
\begin{gather}
  \label{edge-cond}
  \qq_{21} = 0
\end{gather}
It is then easy to verify that
\begin{gather}
  \label{edge-wf-pair}
  \psi_{\ell q+1} = \qqe{11}^\ell \psi_1, \quad 
  \psi_{\ell q} = 0
\end{gather}
where $\ell = 1, 2, \ldots,L$, hence the solution is an edge state
exponentially localized at $y=N_y$ if $|\qq_{11}| > 1$, and at $y=1$
if $|\qq_{11}|< 1$.

The edge spectrum $\{\varepsilon\}$ satisfying the condition
$\qq_{21}(\varepsilon) = 0$ is the same as the \emph{full} spectrum of
a $(q-1) \times (q-1)$ system, so numerically the edge spectrum of
$H(k_x, N_y=Lq-1, z=0)$ can be solved by diagonalizing its
upper-left $(q-1) \times (q-1)$ submatrix.

Note that eqn.~\ref{edge-wf-pair} implies the edge states, \emph{with
  $\psi_0$ included}, has a direct product form
\begin{gather}
  |\psi\rangle =
  \begin{pmatrix}
    \qq_{11}^0\\ \qq_{11}^1 \\ \qq_{11}^2 \\ \vdots \\ \qq_{11}^{L-1}
  \end{pmatrix}\otimes
  \begin{pmatrix}
    \psi_0\\\psi_1\\\psi_2\\ \vdots \\ \psi_{q-1}
  \end{pmatrix}\ ,
\end{gather}
\emph{i.e.}, $\psi_{\ell q + m} = \qq_{11}^{\ell} \psi_m$ with $\ell =
0, 1, \cdots L-1$ and $m = 0, 1, \cdots q-1$. The $N$-component magnetic cell
part dictates the real-space behavior. In this case it is
exponentially localized at either end. The $q$-component internal part
is obtained by prepending $\psi_0 = 0$ to the solutions of the
$(q-1)\times (q-1)$ upper-left block of $H$. This is by no means a
general form of edge states, but we do also notice a similar
decomposition in the zigzag edge modes of the Haldane
model\cite{ha12-mblh}. Note also that all Bloch states have such a
decomposition, $|\Psi(k,n)\rangle = |k\rangle \otimes
|\psi_n(k)\rangle$ where $\langle y | k\rangle = e^{i k y}/\sqrt{N}$
is the Bloch phase and $|\psi_n(k)\rangle$ is the $n^{th}$ band
eigenstate of the Fourier transformed $q\times q$ Hamiltonian. One may
then say that $-i\log(\qq_{11})$ is the imaginary Bloch vector, and
which of the UHP or LHP it resides in tells the localization of the
edge states.

\subsection{Incommensurate edge spectrum}

In the thermodynamic limit where $N_y \rightarrow \infty$, one can
extend the commensurate argument to incommensurate cases, $N_y + 1
= L q + m$,  with $m = 0,1,2, \ldots, q-1$.

First, we note two properties of the transfer matrix,
\begin{gather}
  \label{det}
  \textsf{det}\,(M_y) = 1\\
  M_{y+m}(k_x,\varepsilon) = M_y(k_x + m\phi, \varepsilon)
  \label{Mshift}
\end{gather}
both are straightforward from definition. Eqn.~\ref{Mshift}
expresses the same $k_x\,y$ covariance as eqn.~\ref{kq}. The $(L q + m)$-step transfer matrix can then be divided in two ways,
\begin{align}
  \mm{L q +m}(k_x) &= \mm{m}(k_x) \qq^L(k_x) \\
  &= \qq^L(k_x + m\phi) \mm{m}(k_x)
\end{align}

If $\qq(k_x)$ satisfies the commensurate edge condition eqn.~\ref{edge-cond}, then
\begin{gather}
  \qq^L(k_x) = \begin{pmatrix}
    \qqe{11}^L & x\\  0 & \qqe{22}^L
  \end{pmatrix}_{k_x} \ ,
\end{gather}
where $x$ is some number of no interest.  We then have
\begin{gather}
  \mm{L q+m}(k_x) =
  \underbrace{\begin{pmatrix}
    A_{11} & A_{12}\\ A_{21} & A_{22}
  \end{pmatrix}}_{\mm{m}(k_x)}
  \begin{pmatrix}
    \qqe{11}^L & x\\ 0 & \qqe{22}^L
  \end{pmatrix}_{k_x}\ ,
\end{gather}
hence
\begin{align}
  \label{breaklow}
  \mme{L q + m}_{21}(k_x) &= A_{21} \qqe{11}^L(k_x) \\
  \notag &=  A_{21}\qqe{22}^{-L}(k_x)
\end{align}
where the second equality follows from $ \textsf{det}\, \qq = 1$, a consequence of eqn.~\ref{det}.

Similarly, if $\qq(k_x + m\phi)$ satisfies the commensurate edge condition eqn.~\ref{edge-cond}, we have instead
\begin{gather}
  \mm{L q + m}(k_x) =
  \begin{pmatrix}
    \qqe{11}^L & x\\  0 & \qqe{22}^L
  \end{pmatrix}_{k_x+m\phi}
  \begin{pmatrix}
    A_{11} & A_{12}\\ A_{21} & A_{22}
  \end{pmatrix}
\end{gather}
thus
\begin{align}
  \label{breakupp}
  \mme{L q + m}_{21}(k_x) &= \qqe{22}^L(k_x + m\phi) A_{21}\\
  \notag &= \qqe{11}^{-L}(k_x + m\phi) A_{21}
\end{align}

We can then conclude that in the limit $L \rightarrow \infty$,
\begin{enumerate}
\item If $(k_x, \varepsilon)$ is a solution of the commensurate case $N_y + 1 = L q$ at the \emph{lower} edge,
  $\qqe{11}^L(k_x)\rightarrow~ 0$, then by eqn.~\ref{breaklow}, it is also a solution of incommensurate $N_y + 1 = Lq+m$.
  For $y \in [1,Lq - 1]$, the wavefunction $\psi_y$ coincides with that of the
  commensurate case, and in the upper tail where $y = Lq+m$, $\psi_y =
  A_{11}\qqe{11}^L \psi_1 \rightarrow 0$, thus it is also at the lower edge.
  
\item If $(k_x, \varepsilon)$ is a solution of $N_y + 1 =q L$ at the
  \emph{upper} edge, $\qqe{11}^{-L}(k_x) \rightarrow~ 0$, then by
  eqn.~\ref{breakupp}, $(k_x-m\phi, \varepsilon)$ will be a solution
  of incommensurate $N_y + 1 = Lq+m$. It is also at the upper edge
  because $\psi_y \rightarrow 0$ in the lower tail.
\end{enumerate}
thus for $N_y \rightarrow \infty$, an increment of $N_y$ by $1$ leaves
the lower edge spectrum unchanged, while shifting the upper edge
spectrum in $k_x$ by $-2\pi p/q$.

\section{Triangular lattice}
\label{triangle}
\begin{figure}[!t]
  \centering
  \includegraphics[width=8cm,height=12cm,angle=0]{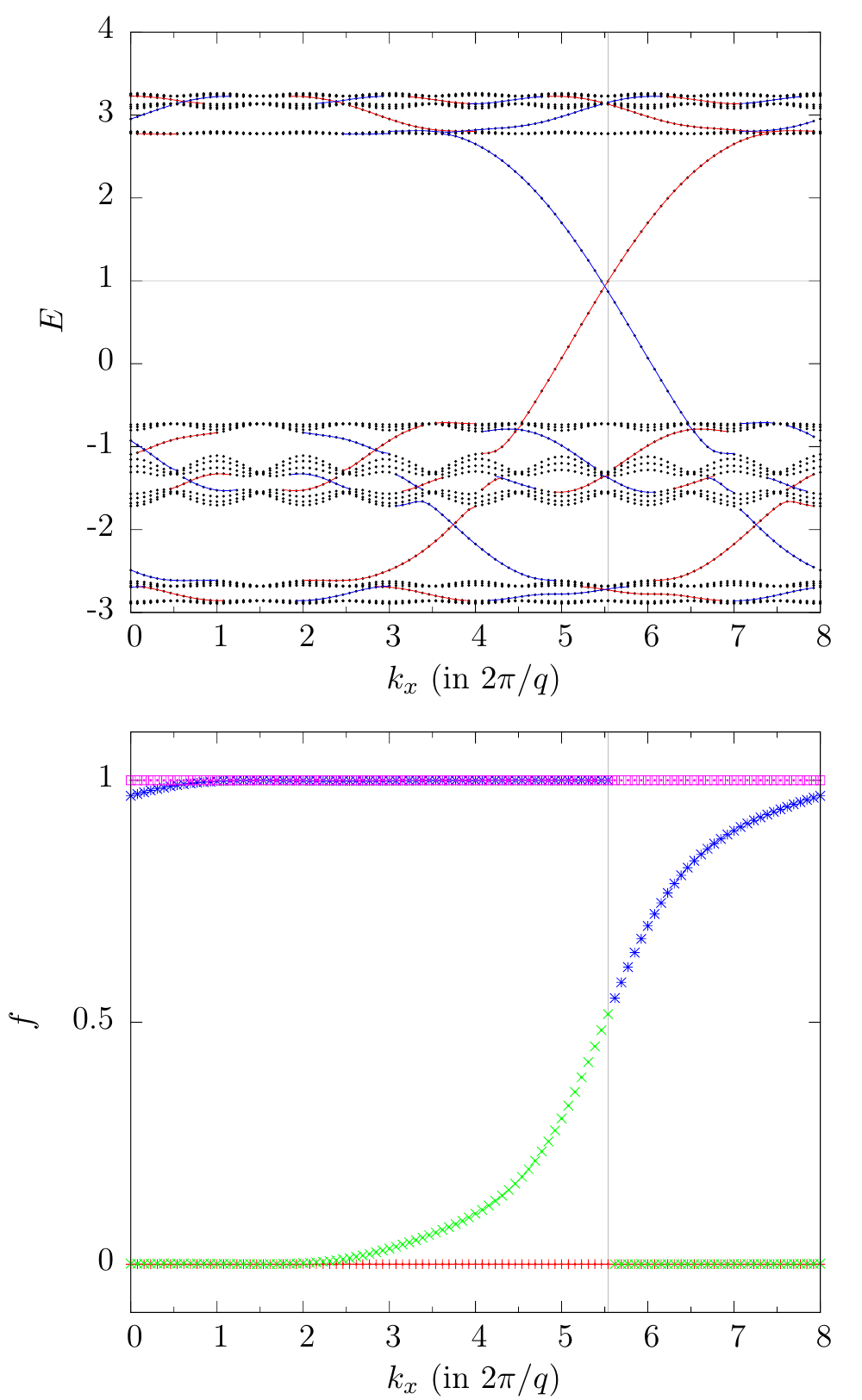}  
  \caption{(Color online) Energy and entanglement spectrum on a
    triangular lattice with cylindrical boundary condition. Parameters used
    are $p/q = 5/8$, $N_y = 32$ and $\EF = 1$. In the energy spectrum
    (top panel), black dots represent bulk levels, red lines represent
    edge modes localized along the lower edge ($y=1$), and blue lines
    represent those localized along the upper edge ($y=N_y = 32$).
    Vertical gray line indicates the $k_x$ value at which $\EF$
    intersects the lower edge state. The entanglement occupancies
    (bottom panel) are computed for the lower half of the cylinder ($1
    \le y \le 16$), color and symbol-coded according to $a$. The
    sudden color change happens when the lower edge mode crosses
    $\EF$.}
  \label{trig-ergent}
\end{figure}
\begin{figure*}[!t]
  \centering
  \includegraphics[width=\textwidth,angle=0]{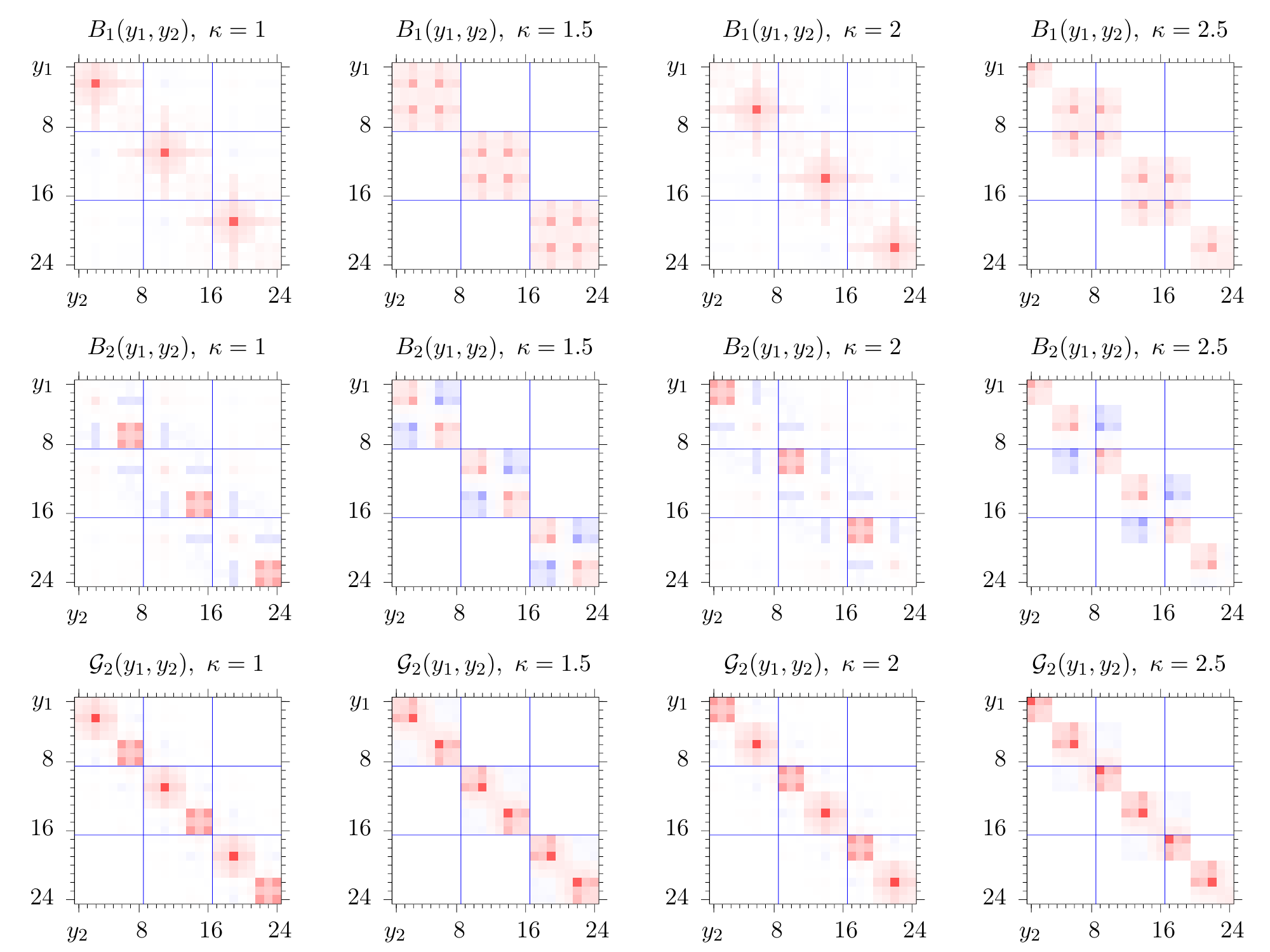}  
  \caption{(Color online) Full system band projectors for $p/q=5/8$,
    $N_y = 32$, with periodic boundary conditions in $y$ and $k_x =
    2\pi \kappa/q$, $B_j$ and $\mathcal{G}_{\nu}$ are projectors for
    the $j^{\rm th}$ band and $\nu$ lowest bands, repectively, see
    also Fig.~\ref{bigfig}. Here we only plot the $24\times 24$
    submatrix belonging to the first three magnetic unit cells. Since
    the Hamiltonian eqn.~\ref{hkx-trig} can no longer be made purely
    real, the projectors are in general complex, so we only plot the
    real part of their matrix elements. Their signs are represented by
    color, red for positive and blue for negative, and their
    magnitudes represented by intensity. The individual bands $B_1$
    and $B_2$ are better localized at $\kappa = 1$ where the gap
    between them is maximal, while their sum, $\mathcal{G}_2$, is
    better localized at $\kappa = 1.5$ where the gap between them is
    minimal. Note that at both $\kappa$, the off-diagonal subblocks of
    both $B_j$ within each $8\times 8$ block tend to cancel (they have
    different colors). The adiabatic evolution of the wave packets are
    obvious: for example, at $\kappa = 1.5$, the wave packets of
    $\mathcal{G}_2$, in each unit cell, are at $y = 3$ and $6$ (mod
    $q$. Same below). At $\kappa = 2$, the wave packet at $y = 3$ is
    in the progress of moving toward $y = 1$ while the one at $y = 6$
    is ``frozen''. At $\kappa = 2.5$, the first wave packet arrives at
    $y = 1$. In the next $\kappa \rightarrow \kappa + 1$ sub-cycle,
    the first wave packet will be frozen and the second one will
    migrate in the diagonal line by $-2$ (negative of the total Chern
    number).}
  \label{trig-proj}
\end{figure*}
The Hofstadter model on a triangular lattice can be obtained by adding
in each square plaquette a diagonal bond along the $\hat x - \hat y$
direction with half-odd-integer vector potential (in units of $\phi =
2\pi p/q$), such that the flux per triangle is $\phi / 2$. Instead of
eqn.~\ref{hkx}, the Hamiltonian matrix is now
\begin{gather}
  \label{hkx-trig}
  H(k_x, N_y, z) = -
  \begin{pmatrix}
    c_1 & v_1 & 0& \cdots & z^{*} v^{*}_{N_y - 1}\\
    v_1^{*} & c_2 & v_2  & & 0\\
    0 & v_2^{*} & \ddots & & \vdots\\
    \vdots & & & &v_{\substack{\\N_y - 1}}\\
    z v_{\substack{\\N_y}} & 0 & \cdots & v_{N_y - 1}^{*} & c_{\substack{\\N_y}}
  \end{pmatrix}
\end{gather}
where
\begin{gather}
  c_y = 2\cos(k_x + y\phi)\quad , \quad v_y = 1 + t' \, e^{-i(k_x + y\phi +
    \frac{1}{2}\phi)}\ ,
\end{gather}
with $t' = 1$ for triangular lattice, and $0$ for square lattice. The
discussion of edge spectrum in Appendix \ref{incommensurate} remains
essentially the same, except the transfer matrix, eqn.~\ref{tmat}, now
becomes
\begin{gather}
  M_y =
  \begin{pmatrix}
    -\dfrac{\varepsilon + c_y}{v_y} & - \dfrac{v^{*}_{y-1}}{v_y}\\
    \\
    1 & 0
  \end{pmatrix}\ .
\end{gather}
In the pathological case where certain $v_{\tilde y} = 0$, the open
edge Hamiltonian (\emph{i.e.}, $z = 0$) reduces to two blocks,
$\{1,\ldots, \tilde y\}$ and $\{\tilde y+1,\ldots, N_y\}$, each of which can be
individually solved; alternatively one can shift $t$ slightly away
from $1$. Note that while $ \textsf{det}\, M_y$ is no longer one, the $q$-step
transfer matrix $\qq$ still has unimodular determinant,
\begin{gather}
  \left|  \textsf{det}\, \qq \right| = \left| \frac{v^{*}_0 v^{*}_1 \cdots
      v^{*}_{q-1}}{ v_1v_2\cdots v_q}\right| = 1\ ,
\end{gather}
where we used $v_0 = v_q$. Consequently,
\begin{gather}
  \left|Q_{22}\right| = \left|Q_{11}\right|^{-1}\ ,
\end{gather}
and eqns.~\ref{breaklow} and \ref{breakupp} hold up to a phase. The
conclusion thus remains unchanged that lower edge states are unchanged
while upper edge states shift in $k_x$ with incommensurate $N_y$.

In Fig.~\ref{trig-ergent}, we plot the cylindrical boundary energy spectrum,
and its entanglement spectrum with $\nu = 5$ filled bulk bands for
$p/q = 5/8$ on the triangular lattice. The Chern numbers of individual
bands are either $C = -3$ or $5$, which are equivalent \emph{modulo}
$q=8$, and the lowest band has $C = -3$. Both are in agreement with
what we observed in the square lattice case, namely, all band Chern
numbers are equivalent \emph{modulo} $q$, and that the lowest band
Chern number is the one with smaller magnitude. The number of edge
spectral flows in each gap is the total band Chern number below the
gap, and the entanglement spectral flow mimics the behavior of edge
spectral flow, and has an index discontinuity at $k_x$ where Fermi
energy intersects the lower edge mode. Note that for $\nu = 5$ filled
bands, the total Chern number is $1$, as reflected in the number of
edge and entanglement spectral flows. This agrees with our observation
that the total Chern number of $p$ filled bands is one, see discussion
following eqn.~\ref{diophantine} in the text.

As in the square lattice case, the band projectors and their sums
also flow under adiabatic $k_x$ pumping, and the number of wave
packets crossing any given boundary during one cycle of the pumping
reflects the Chern number of the projectors. Fig.~\ref{trig-proj}
shows the flow of the lowest two band projectors, $B_1$ and $B_2$, and
their sum $\mathcal{G}_2$, at $k = \kappa\cdot 2\pi/ q$ with $\kappa =
1$ and $1.5$. Both $B_j$ and $\mathcal{G}_{\nu}$ have better
localization at either integer or half-odd-integer $\kappa$ where its
gap from neighboring bands are maximal.

We thus conclude that the observations as detailed in the text using
square lattice are robust and insensitive to the underlying lattice
used.

\bibliographystyle{apsrev-no-url}
\bibliography{hof}

\begin{thebibliography}{26}
\expandafter\ifx\csname natexlab\endcsname\relax\def\natexlab#1{#1}\fi
\expandafter\ifx\csname bibnamefont\endcsname\relax
  \def\bibnamefont#1{#1}\fi
\expandafter\ifx\csname bibfnamefont\endcsname\relax
  \def\bibfnamefont#1{#1}\fi
\expandafter\ifx\csname citenamefont\endcsname\relax
  \def\citenamefont#1{#1}\fi
\expandafter\ifx\csname url\endcsname\relax
  \def\url#1{\texttt{#1}}\fi
\expandafter\ifx\csname urlprefix\endcsname\relax\def\urlprefix{URL }\fi
\providecommand{\bibinfo}[2]{#2}
\providecommand{\eprint}[2][]{\url{#2}}

\bibitem[{\citenamefont{Thouless et~al.}(1982)\citenamefont{Thouless, Kohmoto,
  Nightingale, and den Nijs}}]{tknn82}
\bibinfo{author}{\bibfnamefont{D.~J.} \bibnamefont{Thouless}},
  \bibinfo{author}{\bibfnamefont{M.}~\bibnamefont{Kohmoto}},
  \bibinfo{author}{\bibfnamefont{M.~P.} \bibnamefont{Nightingale}},
  \bibnamefont{and} \bibinfo{author}{\bibfnamefont{M.}~\bibnamefont{den Nijs}},
  \bibinfo{journal}{Phys. Rev. Lett.} \textbf{\bibinfo{volume}{49}},
  \bibinfo{pages}{405} (\bibinfo{year}{1982}).

\bibitem[{\citenamefont{Hofstadter}(1976)}]{hof76}
\bibinfo{author}{\bibfnamefont{D.~R.} \bibnamefont{Hofstadter}},
  \bibinfo{journal}{Phys. Rev. B} \textbf{\bibinfo{volume}{14}},
  \bibinfo{pages}{2239} (\bibinfo{year}{1976}).

\bibitem[{\citenamefont{Hatsugai}(1993)}]{hatsugai93}
\bibinfo{author}{\bibfnamefont{Y.}~\bibnamefont{Hatsugai}},
  \bibinfo{journal}{Phys. Rev. Lett.} \textbf{\bibinfo{volume}{71}},
  \bibinfo{pages}{3697} (\bibinfo{year}{1993}).

\bibitem[{\citenamefont{Li and Haldane}(2008)}]{Li-Haldane08}
\bibinfo{author}{\bibfnamefont{H.}~\bibnamefont{Li}} \bibnamefont{and}
  \bibinfo{author}{\bibfnamefont{F.~D.~M.} \bibnamefont{Haldane}},
  \bibinfo{journal}{Phys. Rev. Lett.} \textbf{\bibinfo{volume}{101}},
  \bibinfo{pages}{010504} (\bibinfo{year}{2008}).

\bibitem[{\citenamefont{Haldane}(2009)}]{haldane09}
\bibinfo{author}{\bibfnamefont{F.~D.~M.} \bibnamefont{Haldane}},
  \bibinfo{journal}{Bull. Am. Phys. Soc.} \textbf{\bibinfo{volume}{54}}
  (\bibinfo{year}{2009}).

\bibitem[{\citenamefont{Cheong and Henley}(2004{\natexlab{a}})}]{cheong04a}
\bibinfo{author}{\bibfnamefont{S.-A.} \bibnamefont{Cheong}} \bibnamefont{and}
  \bibinfo{author}{\bibfnamefont{C.~L.} \bibnamefont{Henley}},
  \bibinfo{journal}{Phys. Rev. B} \textbf{\bibinfo{volume}{69}},
  \bibinfo{pages}{075111} (\bibinfo{year}{2004}{\natexlab{a}}).

\bibitem[{\citenamefont{Cheong and Henley}(2004{\natexlab{b}})}]{cheong04b}
\bibinfo{author}{\bibfnamefont{S.-A.} \bibnamefont{Cheong}} \bibnamefont{and}
  \bibinfo{author}{\bibfnamefont{C.~L.} \bibnamefont{Henley}},
  \bibinfo{journal}{Phys. Rev. B} \textbf{\bibinfo{volume}{69}},
  \bibinfo{pages}{075112} (\bibinfo{year}{2004}{\natexlab{b}}).

\bibitem[{\citenamefont{Peschel}(2003)}]{peschel03}
\bibinfo{author}{\bibfnamefont{I.}~\bibnamefont{Peschel}}, \bibinfo{journal}{J.
  Phys. A} \textbf{\bibinfo{volume}{36}}, \bibinfo{pages}{L205}
  (\bibinfo{year}{2003}).

\bibitem[{\citenamefont{Peschel}(2004)}]{peschel04}
\bibinfo{author}{\bibfnamefont{I.}~\bibnamefont{Peschel}}, \bibinfo{journal}{J.
  Stat. Mech.} \textbf{\bibinfo{volume}{2004}}, \bibinfo{pages}{P06004}
  (\bibinfo{year}{2004}).

\bibitem[{\citenamefont{Turner et~al.}(2010)\citenamefont{Turner, Zhang, and
  Vishwanath}}]{turner10-inversion}
\bibinfo{author}{\bibfnamefont{A.~M.} \bibnamefont{Turner}},
  \bibinfo{author}{\bibfnamefont{Y.}~\bibnamefont{Zhang}}, \bibnamefont{and}
  \bibinfo{author}{\bibfnamefont{A.}~\bibnamefont{Vishwanath}},
  \bibinfo{journal}{Phys. Rev. B} \textbf{\bibinfo{volume}{82}},
  \bibinfo{pages}{241102} (\bibinfo{year}{2010}).

\bibitem[{\citenamefont{Hughes et~al.}(2011)\citenamefont{Hughes, Prodan, and
  Bernevig}}]{hughes-prodan-bernevig11-inversion}
\bibinfo{author}{\bibfnamefont{T.~L.} \bibnamefont{Hughes}},
  \bibinfo{author}{\bibfnamefont{E.}~\bibnamefont{Prodan}}, \bibnamefont{and}
  \bibinfo{author}{\bibfnamefont{B.~A.} \bibnamefont{Bernevig}},
  \bibinfo{journal}{Phys. Rev. B} \textbf{\bibinfo{volume}{83}},
  \bibinfo{pages}{245132} (\bibinfo{year}{2011}).

\bibitem[{\citenamefont{Qi et~al.}(2006)\citenamefont{Qi, Wu, and
  Zhang}}]{qi-wu-zhang06-tbc}
\bibinfo{author}{\bibfnamefont{X.-L.} \bibnamefont{Qi}},
  \bibinfo{author}{\bibfnamefont{Y.-S.} \bibnamefont{Wu}}, \bibnamefont{and}
  \bibinfo{author}{\bibfnamefont{S.-C.} \bibnamefont{Zhang}},
  \bibinfo{journal}{Phys. Rev. B} \textbf{\bibinfo{volume}{74}},
  \bibinfo{pages}{045125} (\bibinfo{year}{2006}).

\bibitem[{\citenamefont{Kohn}(1959)}]{kohn59-wannier}
\bibinfo{author}{\bibfnamefont{W.}~\bibnamefont{Kohn}}, \bibinfo{journal}{Phys.
  Rev.} \textbf{\bibinfo{volume}{115}}, \bibinfo{pages}{809}
  (\bibinfo{year}{1959}).

\bibitem[{\citenamefont{Soluyanov and
  Vanderbilt}(2011)}]{soluyanov-vanderbilt11-z2-wannier}
\bibinfo{author}{\bibfnamefont{A.~A.} \bibnamefont{Soluyanov}}
  \bibnamefont{and}
  \bibinfo{author}{\bibfnamefont{D.}~\bibnamefont{Vanderbilt}},
  \bibinfo{journal}{Phys. Rev. B} \textbf{\bibinfo{volume}{83}},
  \bibinfo{pages}{035108} (\bibinfo{year}{2011}).

\bibitem[{\citenamefont{Yu et~al.}(2011)\citenamefont{Yu, Qi, Bernevig, Fang,
  and Dai}}]{Yu11-z2-wannier}
\bibinfo{author}{\bibfnamefont{R.}~\bibnamefont{Yu}},
  \bibinfo{author}{\bibfnamefont{X.~L.} \bibnamefont{Qi}},
  \bibinfo{author}{\bibfnamefont{A.}~\bibnamefont{Bernevig}},
  \bibinfo{author}{\bibfnamefont{Z.}~\bibnamefont{Fang}}, \bibnamefont{and}
  \bibinfo{author}{\bibfnamefont{X.}~\bibnamefont{Dai}},
  \bibinfo{journal}{Phys. Rev. B} \textbf{\bibinfo{volume}{84}},
  \bibinfo{pages}{075119} (\bibinfo{year}{2011}).

\bibitem[{\citenamefont{Brouder et~al.}(2007)\citenamefont{Brouder, Panati,
  Calandra, Mourougane, and Marzari}}]{brouder07-wannier-obstruction}
\bibinfo{author}{\bibfnamefont{C.}~\bibnamefont{Brouder}},
  \bibinfo{author}{\bibfnamefont{G.}~\bibnamefont{Panati}},
  \bibinfo{author}{\bibfnamefont{M.}~\bibnamefont{Calandra}},
  \bibinfo{author}{\bibfnamefont{C.}~\bibnamefont{Mourougane}},
  \bibnamefont{and} \bibinfo{author}{\bibfnamefont{N.}~\bibnamefont{Marzari}},
  \bibinfo{journal}{Phys. Rev. Lett.} \textbf{\bibinfo{volume}{98}},
  \bibinfo{pages}{046402} (\bibinfo{year}{2007}).

\bibitem[{\citenamefont{Rodr{\'\i}guez and Sierra}(2009)}]{Rodriguez09}
\bibinfo{author}{\bibfnamefont{I.~D.} \bibnamefont{Rodr{\'\i}guez}}
  \bibnamefont{and} \bibinfo{author}{\bibfnamefont{G.}~\bibnamefont{Sierra}},
  \bibinfo{journal}{Phys. Rev. B} \textbf{\bibinfo{volume}{80}},
  \bibinfo{pages}{153303} (\bibinfo{year}{2009}).

\bibitem[{\citenamefont{Alexandradinata
  et~al.}(2011)\citenamefont{Alexandradinata, Hughes, and Bernevig}}]{AHB11}
\bibinfo{author}{\bibfnamefont{A.}~\bibnamefont{Alexandradinata}},
  \bibinfo{author}{\bibfnamefont{T.~L.} \bibnamefont{Hughes}},
  \bibnamefont{and} \bibinfo{author}{\bibfnamefont{B.~A.}
  \bibnamefont{Bernevig}}, \bibinfo{journal}{Phys. Rev. B}
  \textbf{\bibinfo{volume}{84}}, \bibinfo{pages}{195103}
  (\bibinfo{year}{2011}).

\bibitem[{\citenamefont{Cloizeaux}(1964)}]{cloizeaux64.paperII}
\bibinfo{author}{\bibfnamefont{J.~D.} \bibnamefont{Cloizeaux}},
  \bibinfo{journal}{Phys. Rev.} \textbf{\bibinfo{volume}{135}},
  \bibinfo{pages}{A685} (\bibinfo{year}{1964}).

\bibitem[{\citenamefont{Thouless}(1984)}]{thouless84}
\bibinfo{author}{\bibfnamefont{D.}~\bibnamefont{Thouless}},
  \bibinfo{journal}{Surface Science} \textbf{\bibinfo{volume}{142}},
  \bibinfo{pages}{147 } (\bibinfo{year}{1984}), ISSN \bibinfo{issn}{0039-6028}.

\bibitem[{\citenamefont{Wen and Zee}(1989)}]{wen89-zm}
\bibinfo{author}{\bibfnamefont{X.}~\bibnamefont{Wen}} \bibnamefont{and}
  \bibinfo{author}{\bibfnamefont{A.}~\bibnamefont{Zee}},
  \bibinfo{journal}{Nuclear Physics B} \textbf{\bibinfo{volume}{316}},
  \bibinfo{pages}{641 } (\bibinfo{year}{1989}), ISSN \bibinfo{issn}{0550-3213}.

\bibitem[{\citenamefont{Kohmoto}(1989)}]{komoto89-zm}
\bibinfo{author}{\bibfnamefont{M.}~\bibnamefont{Kohmoto}},
  \bibinfo{journal}{Phys. Rev. B} \textbf{\bibinfo{volume}{39}},
  \bibinfo{pages}{11943} (\bibinfo{year}{1989}).

\bibitem[{\citenamefont{Qi}(2011)}]{Qi11-wannier}
\bibinfo{author}{\bibfnamefont{X.-L.} \bibnamefont{Qi}},
  \bibinfo{journal}{Phys. Rev. Lett.} \textbf{\bibinfo{volume}{107}},
  \bibinfo{pages}{126803} (\bibinfo{year}{2011}).

\bibitem[{\citenamefont{Provost and Vallee}(1980)}]{provost80}
\bibinfo{author}{\bibfnamefont{J.}~\bibnamefont{Provost}} \bibnamefont{and}
  \bibinfo{author}{\bibfnamefont{G.}~\bibnamefont{Vallee}},
  \bibinfo{journal}{Commun. Math. Phys.} \textbf{\bibinfo{volume}{76}},
  \bibinfo{pages}{289} (\bibinfo{year}{1980}).

\bibitem[{\citenamefont{Campos~Venuti and Zanardi}(2007)}]{zanardi07}
\bibinfo{author}{\bibfnamefont{L.}~\bibnamefont{Campos~Venuti}}
  \bibnamefont{and} \bibinfo{author}{\bibfnamefont{P.}~\bibnamefont{Zanardi}},
  \bibinfo{journal}{Phys. Rev. Lett.} \textbf{\bibinfo{volume}{99}},
  \bibinfo{pages}{095701} (\bibinfo{year}{2007}).

\bibitem[{\citenamefont{{Huang} and {Arovas}}(2012)}]{ha12-mblh}
\bibinfo{author}{\bibfnamefont{Z.}~\bibnamefont{{Huang}}} \bibnamefont{and}
  \bibinfo{author}{\bibfnamefont{D.~P.} \bibnamefont{{Arovas}}},
  \bibinfo{journal}{ArXiv e-prints}  (\bibinfo{year}{2012}),
  \eprint{1205.6266}.

\end{thebibliography}
\end{document}